\documentclass{JINST}

\title{The Planck/LFI Radiometer Electronics Box Assembly}

\author{
J.~M.~Herreros$^a$, 
M.~F.~G\'omez$^a$,
R.~Rebolo$^{a,b}$\thanks{Corresponding
author.},
H.~Chulani$^a$,
J.~A.~Rubi\~no-Martin$^a$,
S.~R.~Hildebrandt$^a$,
M.~Bersanelli$^c$
R.~C.~Butler$^d$,
M.~Miccolis$^e$,
A.~Pe\~na$^f$,
M.~Pereira$^f$,  
F.~Torrero$^f$, 
C.~Franceschet$^c$,
M.~L\'opez$^f$ and
C.~Alcal\'a$^f$ \\
\llap{$^a$} Instituto de Astrof\'isica de Canarias (IAC),\\ 
38200 La Laguna. Tenerife. Spain\\
\llap{$^b$} Consejo Superior de Investigaciones Cient\'ificas, Spain\\
\llap{$^c$} Universit\' a di Milano, Dipartamento di Fisica, \\
Via G. Celoria 16,I-20133 Milano, Italy\\
\llap{$^d$} INAF-IASF Bologna,\\
Via P. Gobetti, 101, I-40129 Bologna, Italy\\
\llap{$^e$} Thales Alenia Space Italia S.p.A.,\\
 IUEL - Scientific Instruments, S.S. Padana Superiore 290, 20090 Vimodrone (Mi)
 Italy\\
\llap{$^f$} EADS Astrium CRISA, Madrid, Spain\\
  E-mail: \email{rrl@iac.es}}

\abstract{ The Radiometer Electronics Box Assembly (REBA) is the control and
  data processing on board computer of the Low Frequency Instrument (LFI) of the
  Planck mission (ESA).
The REBA was designed and built incorporating state of the art processors,
communication interfaces and real time operating system software in order to
meet the scientific performance of the LFI.
We present a technical summary of the REBA, including a physical, functional,
electrical, mechanical and thermal description.
Aspects of the design and development, the assembly, the integration and the
verification of the equipment are provided.  A brief description of the LFI on
board software is given including the Low-Level Software and the main
functionalities and architecture of the Application Software. The compressor
module, which has been developed as an independent product, later integrated in
the application, is also described in this paper.
Two identical engineering models EM and AVM, the engineering qualification model
EQM, the flight model FM and flight spare have been manufactured and
tested. Low-level and Application software have been developed.
Verification activities demonstrated that the REBA hardware and software fulfil
all the specifications and perform as required for flight operation. }

\keywords{Cosmic Microwave Background -- space mission -- Data Handling -
  Digital Signal Processing - Data Compression -- Data Acquisition - Low-Level
  Software -- SpaceWire - MIL-STD-1553B}

\begin{document}

%
\section{Introduction}

The Low Frequency Instrument (LFI, Bersanelli et al.~2009; Mandolesi et
al.~2009) of the ESA satellite Planck consists of an array of 11 corrugated
horns feeding 22 polarisation sensitive pseudo-correlation radiometers based on
HEMT transistors and MMIC technology. These are actively cooled down to 20~K by
using a new concept sorption cooler. The radiometers cover three frequency bands
centred at 30~GHz, 44~GHz, and 70~GHz. The LFI shares the focal plane of the
1.5m-aperture Planck telescope with the High Frequency Instrument (HFI), which
is based on bolometric detectors cooled to 0.1~K (Lamarre et al.~2009). The
combination of LFI and HFI is designed to provide full-sky maps of the microwave
sky with an unprecedented combination of spectral coverage (30-850~GHz),
sensitivity ($\Delta T/T \sim 2\times 10^{-6}$), angular resolution ($\sim
10$~arcmin) and suppression of systematic effects.

Each LFI radiometer correlates the signal from the sky with that of a reference
blackbody source cooled to 4~K. At the end of the radiometric chain, the
amplified signals are detected by square law diodes, DC amplified and
transmitted to the Data Acquisition Electronics (DAE) for signal conditioning
and digitization.

The LFI array required a dedicated electronic system to perform on-board
instrument control, command and data handling as well as the on-board scientific
data processing with high reliability and robustness. These functions are
carried out by the LFI Radiometer Electronics Box Assembly (REBA), which is
described in detail in this paper. In particular, the REBA incorporates the
compression algorithms needed to optimize the use of the satellite to ground
available communication bandwidth.
The expected output of LFI is 5.7~Mbps while the available effective bandwidth
is the order of 53~kbps, implying a need for on board pre-processing of the raw
data. The most relevant physical, functional, electrical, mechanical and thermal
characteristics of the REBA, as well as the main software aspects, are described
in detail in the following sections. For a broad description of the REBA in the
context of LFI, see Section 4.4.2 in Bersanelli et al.~(2009) on the DAE and its
connection with REBA.

%
%

\section{Equipment overview}

\begin{figure*}
\centering
\includegraphics[width=0.35\textwidth]{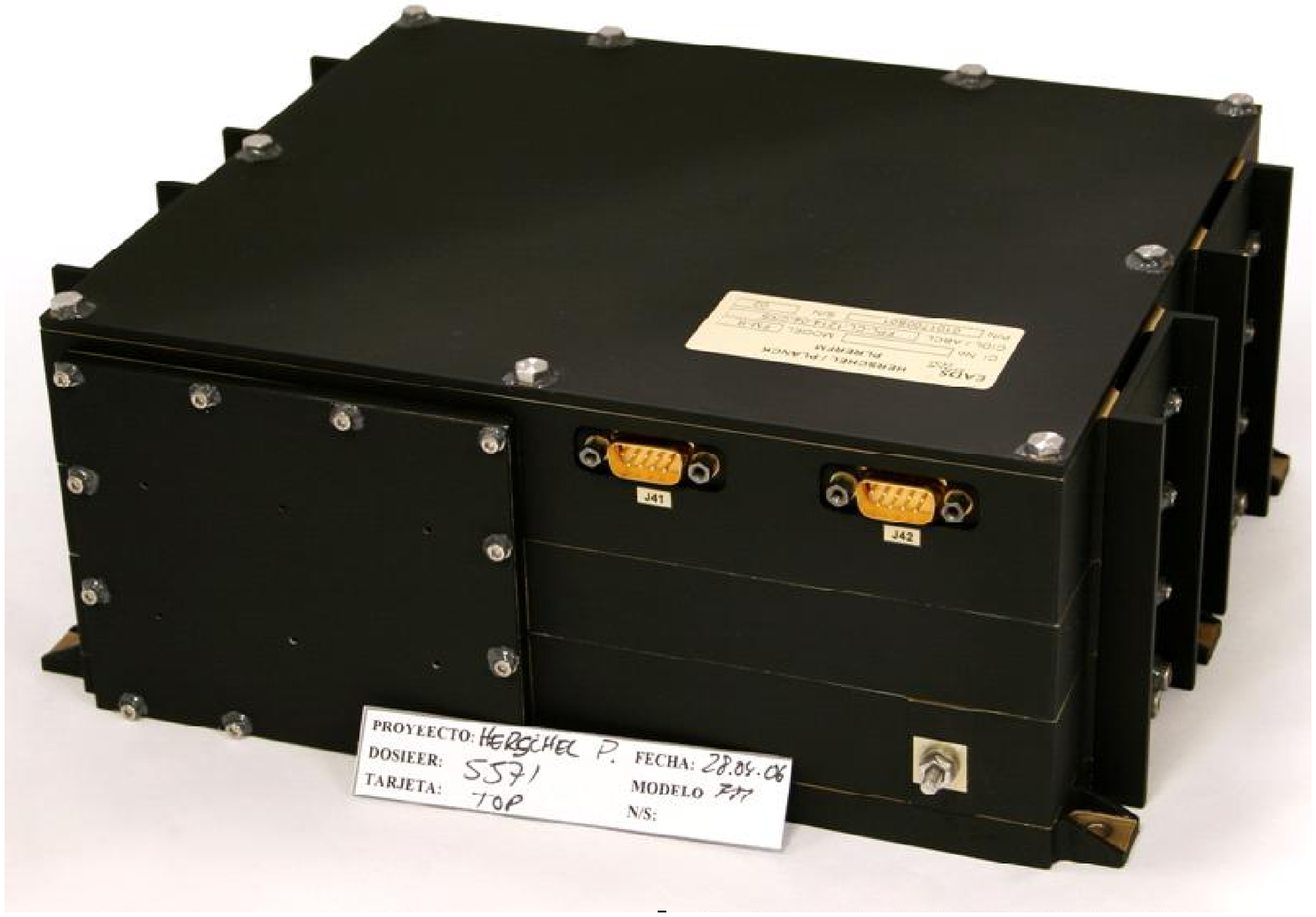}%
\includegraphics[width=0.35\textwidth]{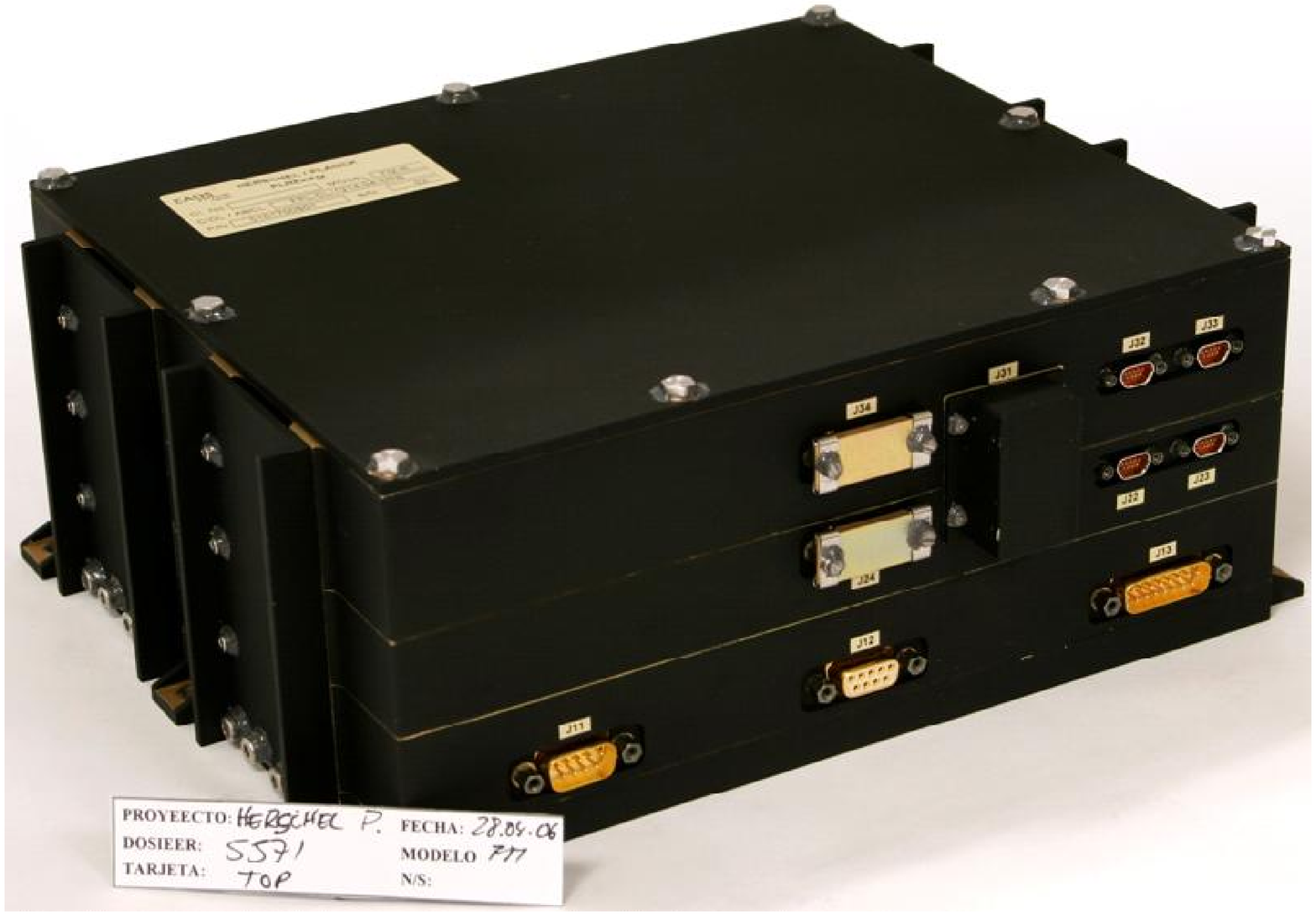}
   \caption{REBA nominal and redundant flight models.}
      \label{fig:1-1}
\end{figure*}

The REBA is a warm electronics unit consisting of two separate identical units,
one nominal and one redundant, which operate in cold redundancy under power
supply control of the spacecraft. Each unit has an envelope of $270 \times 233
\times 108$~mm$^3$, is painted in mat black with an emissivity of 0.9, and
weights 4.32~Kg (see Figure~\ref{fig:1-1}).  The measured power consumption of
the REBA is 22.7~W fully running and processing science data, while the measured
total power consumption of LFI is 68.3~W. Each unit houses three stacked
electronics modules and they are located on the lateral panels of the service
module (SVM) of the spacecraft at about 300~K.

\begin{figure}
\centering
\includegraphics[width=0.85\columnwidth]{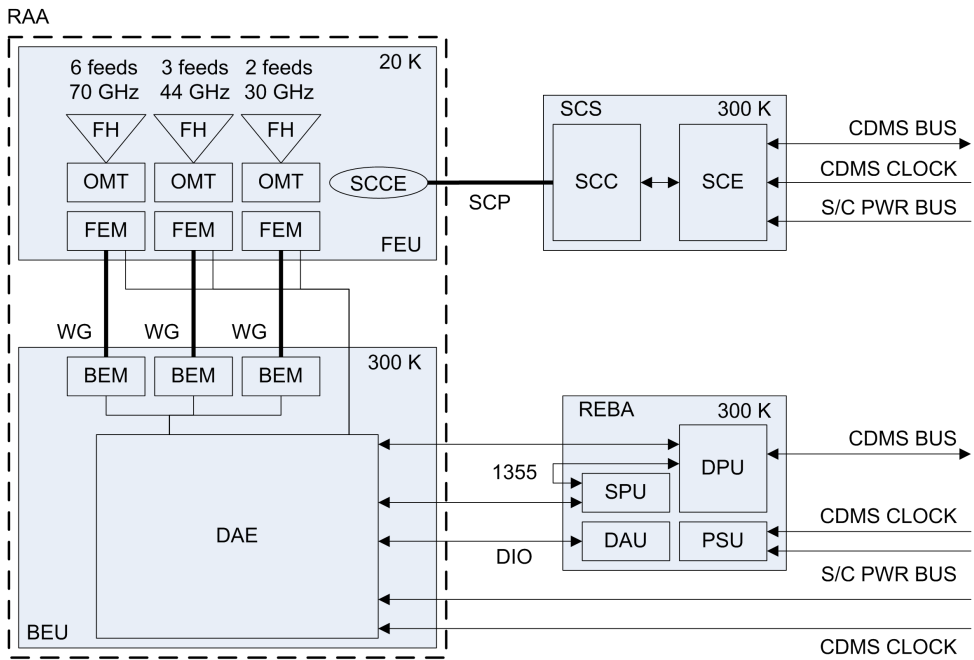}
  \caption{LFI functional block diagram (nominal). }
      \label{fig:1-2}
\end{figure}

Each REBA unit is connected to the instrument Data Adquisition Electronics (DAE)
in the Radiometer Array Assembly (RAA) through the instrument harness and to the
spacecraft Central Data Management Unit (CDMU) and Power Control Distribution
Unit (PCDU) through the spacecraft harness. Figure~\ref{fig:1-2} shows the
connections between the REBA nominal unit, the DAE and the spacecraft CDMU and
PCDU.

The REBA incorporates very innovative advances in several areas, remarkably: i)
new DSP processors like TSC21020E which were considered among the most advanced
high-performance processors (18~MHz and 0 wait state - EDAC protected) for
scientific application in space missions; ii) very high velocity serial
communications interfaces and, iii) new real time on-board operating system.

The REBA provides the hardware to perform the following main functions:
\begin{itemize}
\item On-board command and data handling.
\item On-board science data processing and compression. 
\item Instrument control. 
\item Communication with the spacecraft CDMU via a MIL-STD-1553B CDMS bus interface. 
\item Communication with the LFI subsystem DAE via four IEEE-1355 full-duplex,
  bi-directional, serial, point-to-point data links.
\item Hardware initialisation and error management. 
\item Memory Error Detection and Correction (EDAC). 
\item Computer watchdog activity control. Internal on-board time management and
  synchronisation with the CDMS. 
\item DAE synchronisation.
\item On-board software storage and processing. 
\item Internal housekeeping data acquisition. 
\item S/C Power conditioning and internal power distribution.
\item Digital Input/Output interface with the DAE for control purposes.
\end{itemize}

\begin{figure}
\centering
\includegraphics[width=0.85\columnwidth]{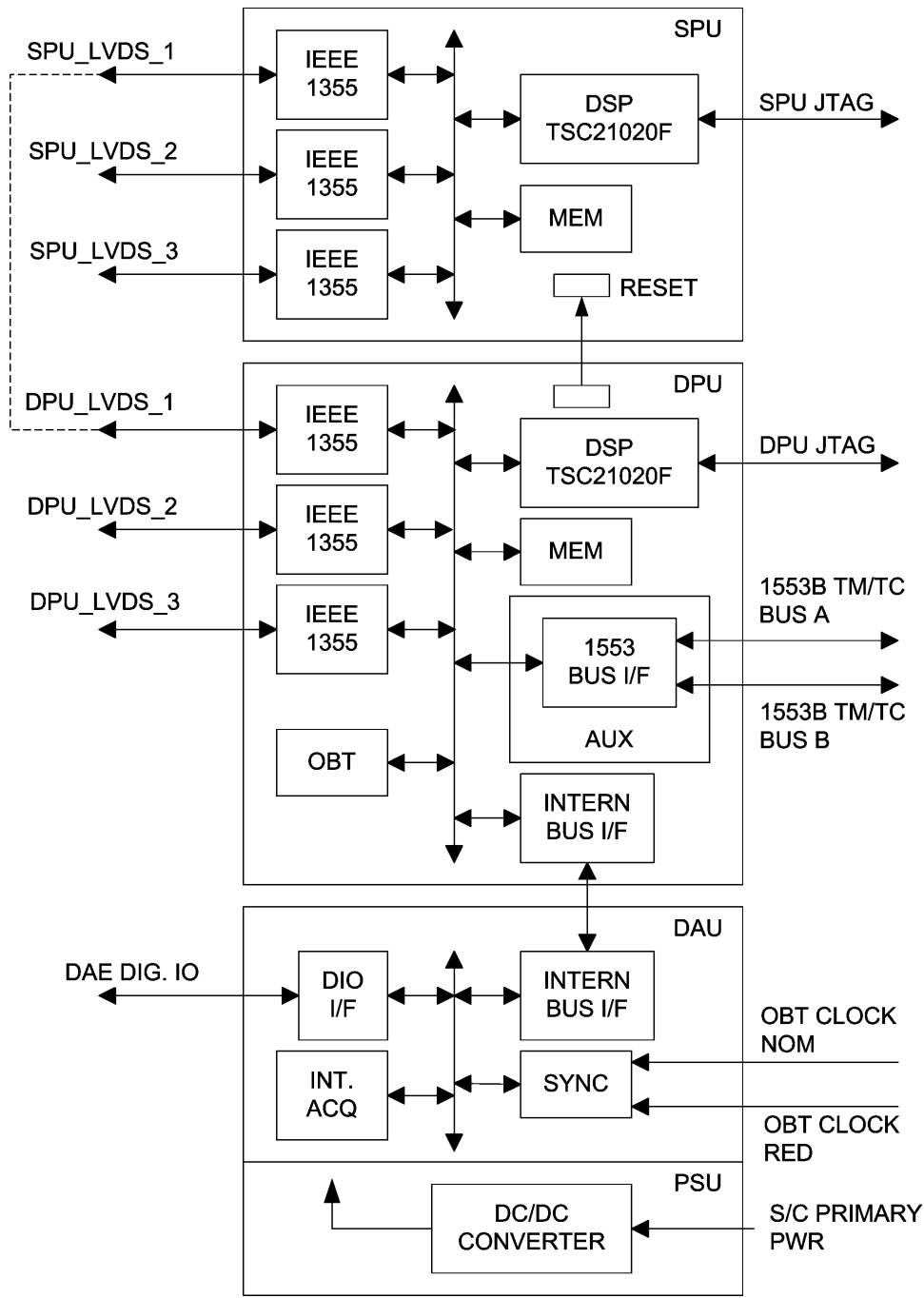}
  \caption{REBA functional block diagram. }
      \label{fig:1-3}
\end{figure}

These functions are allocated to four functional units: the Data Processing Unit
(DPU), the Signal processing Unit (SPU), the Data Acquisition Unit (DAU), and
the Power supply Unit (PSU). The DPU and SPU are in charge of the telecommand,
housekeeping and science telemetry processing and provide the functions to
ensure that the hardware and software operate as planned. The DAU performs the
analogue to digital conversion of the housekeeping of the REBA itself, while the
switching regulator of the PSU provides galvanic isolation converting the
spacecraft (S/C) power bus voltage to a series of current limited, under/over
voltage protected regulated voltages for the sections of the REBA. It performs
the control of the DC/DC converter switching and distributes the electrical
power to the SPU sub-units (DPU, SPU and DAU). Figure~\ref{fig:1-3} shows a
simplified functional block diagram of the REBA.

The REBA  is composed of three boards, two DSP boards and one board for
the DAE interface function and the DC/DC Converter. The DPU Digital Signal Processor (DSP)
 board is in charge of the control of the different
elements: its software controls the DAU board by means of the Internal Bus and
it also controls the SPU DSP board by an external spacewire connection. The DC/DC
converter supplies all the functions. Table~\ref{table:1.1} shows the DPU and
SPU characteristics and performance.

\begin{table}
\scriptsize
\caption{DPU and SPU characteristics and performances. }
\label{table:1.1}     
\centering                
\begin{tabular}{l p{2.5in} l}
\hline\hline
Feature & DPU &  SPU \\          
\hline                      
Program ROM Size	& 32 KW x 48	& 32 KW x 48 \\ 
Program RAM Size	& 512 KW x 48 (0 wait state - EDAC protected) &	512 KW x 48 (0 wait state - EDAC protected) \\
Data RAM Size	        & 512 KW x 32 (0 wait state - EDAC protected) & 	512 KW x 32 (0 wait state - EDAC protected) \\
Exp. Data RAM Size	& Not implemented	& 512 KW x 32 (0 wait state - EDAC protected) \\
EEPROM Size	        & 256 KW x 48	& Not implemented \\
Interface to MIL BUS & Included in Auxiliary Board & Not implemented \\
OBT timer	        & PMPSC	& PMPSC (Not used) \\
Watchdog timer	& DMPSC	& DMPSC \\
SMCS Interface	& \multicolumn{2}{c}{One IF with three LVDS channels}  \\
DSP Operation	        & \multicolumn{2}{c}{18 MHz}  \\
Interface to DAU	& - End of Acquisition Interrupt

- Internal analog acquisitions

- 1Hz Interrupt

- Converter Sync status input

- 1Hz signal control	& Not implemented \\
Interface to DAE	& 
- DAE Power Status input

- DAE reset SMCS1 output

- DAE reset SMCS2 output & 
- DAE Data ready Interrupt \\
Interface to SPU	& - Reset to SPU &	- Reset from DPU \\
\hline    
\end{tabular}
\end{table}

\subsection{DSP module}

\begin{figure}
\centering
\includegraphics[width=0.85\columnwidth]{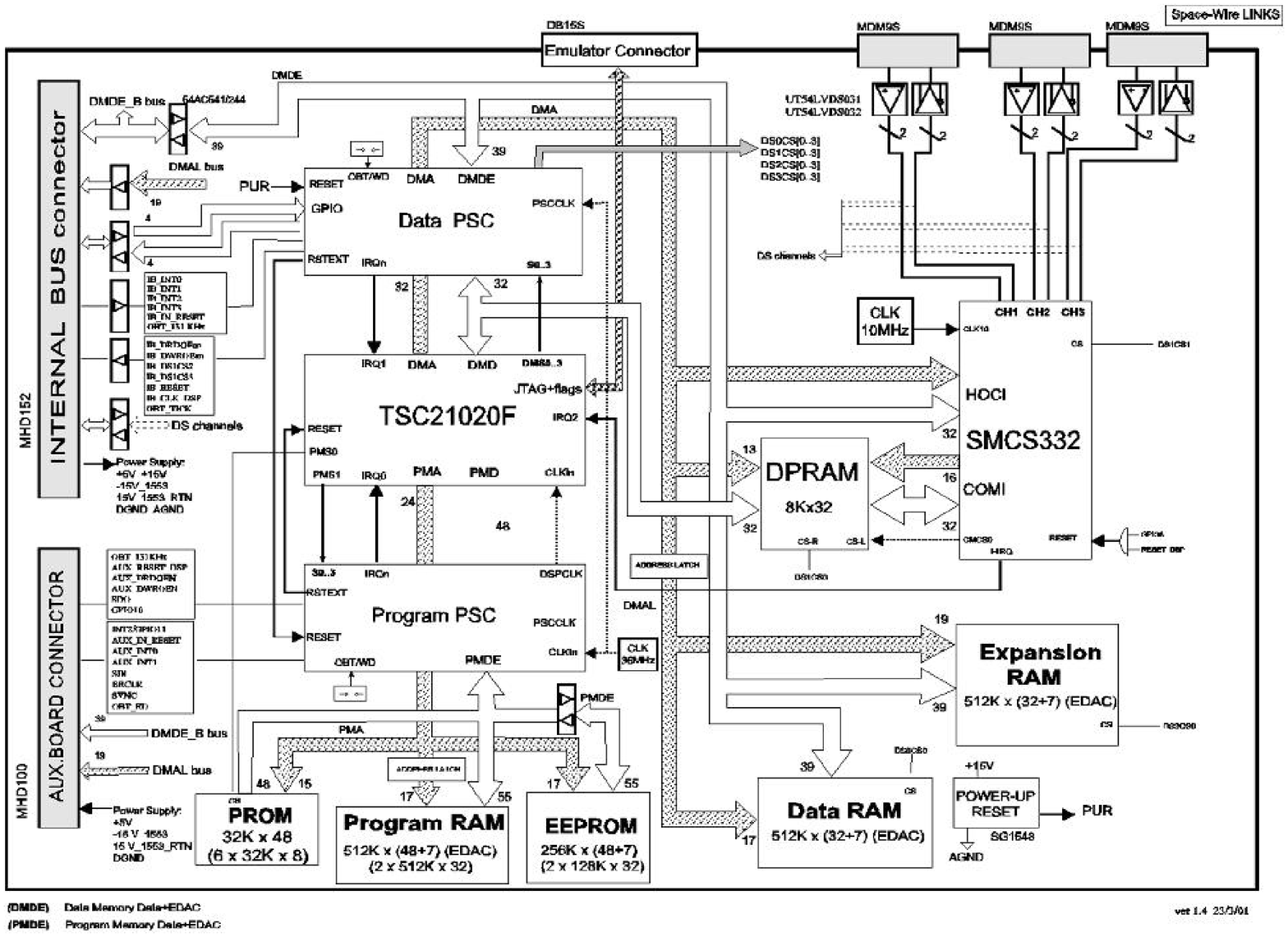}
  \caption{DSP board diagram. }
      \label{fig:1-4}
\end{figure}

The Processor Module - see block diagram in Figure~\ref{fig:1-4}- is a Floating
Point Digital Signal Processing CPU on a single board based on the DSP
TSC21020F, radiation tolerant version of the ADSP21020 from Analogue Devices. Its
external and open Harvard architecture provides two complete bus systems for
program (48 bits) and data (32 bits), allowing concurrent access and
simultaneous fetching and data accesses or multiple data accesses on a single
clock cycle. Expansion capabilities are provided via the backpanel system bus as
well as a mezzanine auxiliary interface. The Processor Module implements into a
single board: 6x32 Kbytes PROM, 256K x 64 bits EEPROM (EDAC protected), 512K x
56 bits program RAM (EDAC protected) and 512K x 40 bits data RAM (EDAC
protected), 512K x 40 bits of Expansion RAM; three high-speed IEEE-1355
(SpaceWire) links, asynchronous serial interface; Expanded interrupt
capabilities; System Bus control; Programmable Watch-dog \& system Timer and
IEEE 1149.1 (JTAG) interface.

General Features: TSC21020F IEEE 32/40 bits Floating Point Digital Signal
Processor at up to 18 MHz, 0 wait states for RAM. Harvard architecture with
independent program and data memory buses. IEEE 1149.1 I/F (JTAG) for testing
and debugging. On-chip emulation. 6x32Kx48bits start-up PROM, 256Kx64bits EEPROM
program bank (EDAC protected), 512Kx56 bits Program RAM (EDAC protected) and
512Kx40 bits Data RAM plus an optional 512Kx40 Expansion RAM bank (both banks
EDAC protected). 6 Interrupt levels (4 of them through the System Bus) and IEEE
Exception Handling with Interrupt on Exception. Bi-directional IEEE-1355
(SpaceWire) based on SMCS332 and LVDS drivers capable of 100 Mbps data
rates. Programmable Watch-Dog and 32 bit system Timers. Power and Reset
monitor. System Bus Interface (Master). Unit Monitoring and Control capabilities
through System Bus. Standard development tools from Analog Devices.

Key Design Features: Performance (at 18 MHz): 18 MIPS, 54 MFLOPS (peak) 36
MFLOPS (sustained). EDAC, UART, Timers, General purpose I/O registers and Glue
logic integrated into the Processor Support Chip (PSC) ASIC designed and
developed by CRISA. High degree of expansion, compatibility and configurability
through the system bus interface and Auxiliary bus interfaces: The System Bus
interface has been designed for connecting the DSP to maximise the expansion
possibilities (I/O modules, etc.). The Auxiliary bus interface has been
designed for connection of a mezzanine auxiliary board originally intended for
high speed interfaces such as the 1553/OBDH bus interface, EEPROM/RAM expansion,
etc.

The purpose of the PSC Chip is: To enhance the on-chip memory management
capabilities of the DSP, providing additional address decoding and wait state
generator. Four IO Areas are provided per decoding bank. To provide a flow
through EDAC, configurable in width and enabling per I/O Area. To provide a
simple 32 bit programmable Timer. To provide a Complex Timer configurable to
provide a System Watchdog or an On-Board Time (OBT) support counter-timer. To
provide a high speed UART. To provide buffer control in order to support the
implementation of extension buses. The design of the PSC is compatible with the
use in the PMB and DMB. It considers the bit aligning and bus wideness.

\subsection{Auxiliary board}

\begin{figure*}
\centering
\includegraphics[width=0.85\textwidth]{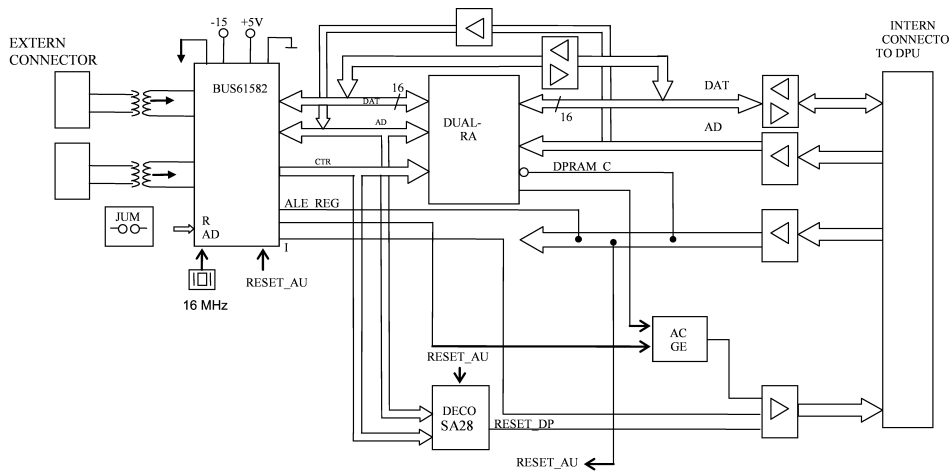}
  \caption{AUX  Board Block Diagram. }
      \label{fig:1-5}
\end{figure*}

The 1553B Interface function is responsible for the communication between the
DPU and the CDMS. As part of the DPU board, this block is implemented on an
independent board inserted in the Auxiliary Connector of the DPU board
itself. The block diagram of this function is shown in Figure~\ref{fig:1-5}. The
external connection is of course to the 1553B Data bus. As required by the
MIL-STD-1553 standard, the connection is made to both buses, and the
implementation is by transformer coupling. The Remote Terminal Address, is
discrepant to the standard being internally hardwired, and is different for the
Nominal and Redudant Units.

The architecture of the 1553B interface is built around the DDC BU61582. This
hardened device provides a completely integrated BC/RT/MT interface between the
host processor and the bus, although its use in the REBA is reduced to Remote
Terminal mode. It integrates dual transceiver, protocol, memory management and
processor interface logic. It also includes internal 16Kwords of RAM which is
not used in the design due to its very low speed characteristics.

The interface for data with the DPU software is performed by means of a
Dual-Port RAM. The software can also access the internal BU61582 internal
registers, and an acknowledge generator is included to allow access even in case
the component is processing commands. This circuit also ensures no conflict
appears in the accesses to the Dual-Port RAM.

The interface also includes circuitry to generate a general DPU reset on
reception of a SA28R command over the bus. The block implements some logic to
avoid feedback problems between this reset line, that is sent directly to the
DPU, and the general reset line from the DPU that is activated immediately
after. It introduces a delay allowing the transmission of the status response
through the 1553 bus after the reception of the command.

The interface with the DPU board through the Auxiliary connector also includes a
buffering stage which is not part of the DSP board. This stage decouples the
internal Aux board operation from the DSP activities. This board uses two
secondary voltages supplies: 5V for the general logic as well as for the BU61582
and $-15$V, a devoted DC/DC converter output with its corresponding return line.
The general logic and DC/DC converter return lines are connected in order to
decouple the large peak currents related to the bus communications from the
general supply lines.

\subsection{DAU function}

The DAU is included on the Power Supply Unit (PSU) board together with the DC/DC
Converter. The DAU acquires the analogue housekeeping of the REBA performing
analogue signal conditioning, digital conversion, and transmission to the DPU
via the back panel connector.

\begin{figure*}
\centering
\includegraphics[width=0.85\textwidth]{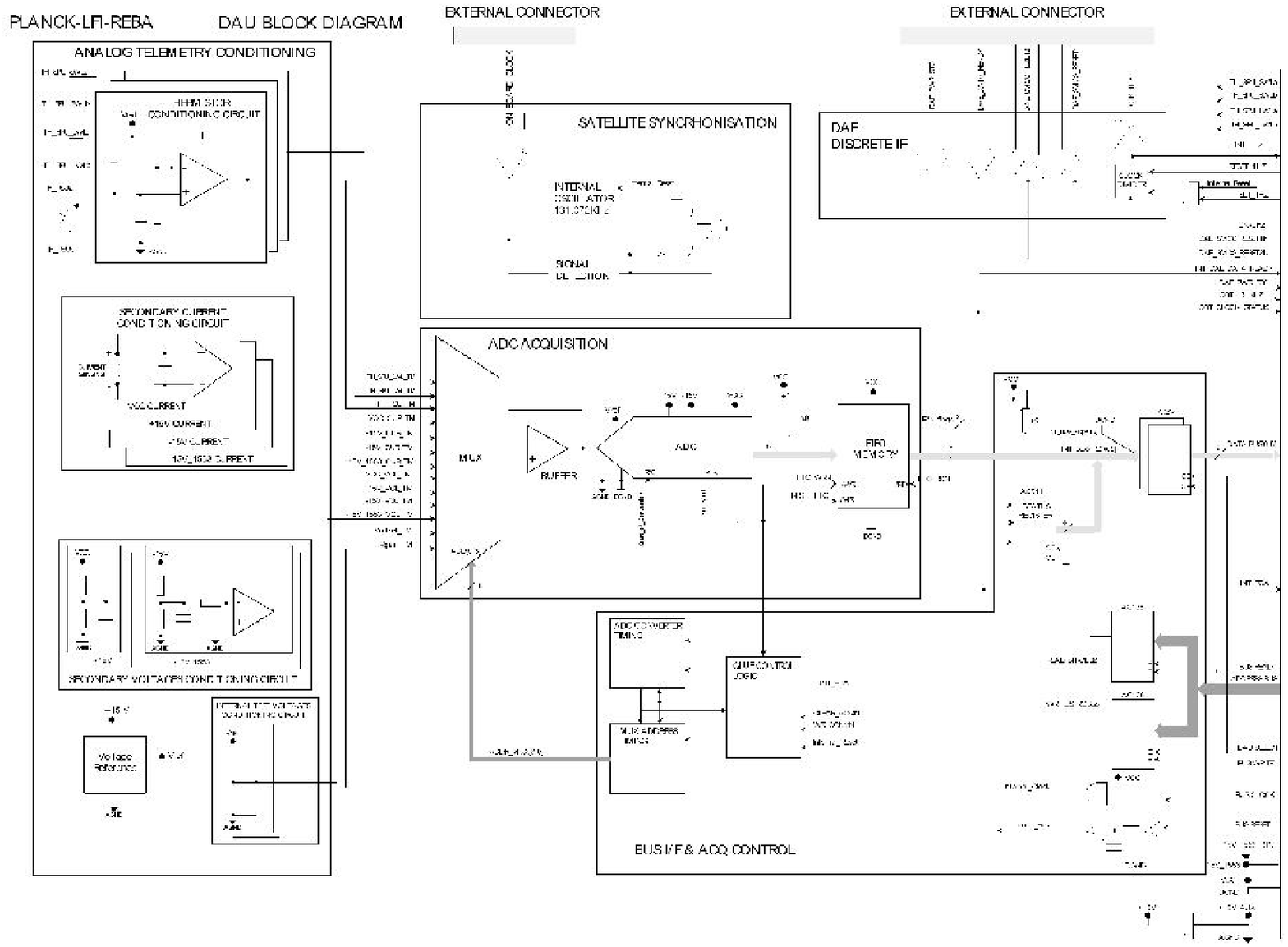}
  \caption{DAU block diagram. }
      \label{fig:1-6}
\end{figure*}

Figure~\ref{fig:1-6} shows the DAU functional blocks. The specific DAU functions
are: Internal Bus Interface. Analogue to Digital Conversion. Acquisition
Control. Clock Divider. Thermistors Conditioning. Secondary Currents
Conditioning. Secondary Voltages Conditioning. Internal Test Voltages
Conditioning. Status.

The DAU is a slave function controlled by the DPU module. The DPU generates the
Bus lines that handle the DAU by means of write and read bus cycles. The "Start
of an acquisition cycle" signal is generated with a write command by the
DPU. This command enables the Acquisition Control Block and starts a cycle of
acquisition. The Status Block contains the following information: ``Acquisition
in progress'' bit, when set, this indicates that an acquisition cycle was started and
has not still finished; when cleared, this indicates that an acquisition cycle
finished and a new cycle can be started. FIFO Flags: Three bits that indicate
the FIFO status (empty, not empty, full, etc.). OBT clock status which indictes
if the REBA OBT counter is using the external or internal clock.

In the Analogue to Digital conversion block the analogue signals are routed to a
16 channel multiplexer stage of type HI-546. The selected signal is buffered by
an OP-42 Operational Amplifier and applied to the input of an Analogue to Digital
Converter (ADC) type AD-574. The 9 most significant bits output of the ADC are
stored in a FIFO memory. The method of acquisition is cyclic. Once all the channels
have been sampled and placed in the FIFO the Acquisition in Progress signal is
deactivated so that the DPU can read out the data.

The RS-422 block of the DAU provides the interface to the DAE for the 1 Hz time
synchronisation signal used by the DAE at power up and during the process of
time synchronisation with the OBT. The signal is derived from the 2 Hz signal
generated by the CPU. Its phase is controlled (by DPU software and by means of
two signals) in order to ensure a rising edge when required, both at power up
and during the synchronization procedure. Three temperature monitoring telemetries
are provided by the DAU function: the DPU board temperature, the SPU board
temperature and the PSU board temperature, using NTC type thermistor. Four
secondary voltage monitoring telemetries are provided by the DAU function: the
VCC voltage (logic supply), the $+15$V voltage (positive analogue supply) and
the $-15$V voltage (negative analogue supply), and the $-15$V\_1553 voltage. The
currents of these secondary voltages, sensed at the active path of the
secondary, are also provided. Each conditioning circuitry includes a filter with
$fc = 100$~Hz.

\subsection{DC/DC Converter}

\begin{figure*}
\centering
\includegraphics[width=0.85\textwidth]{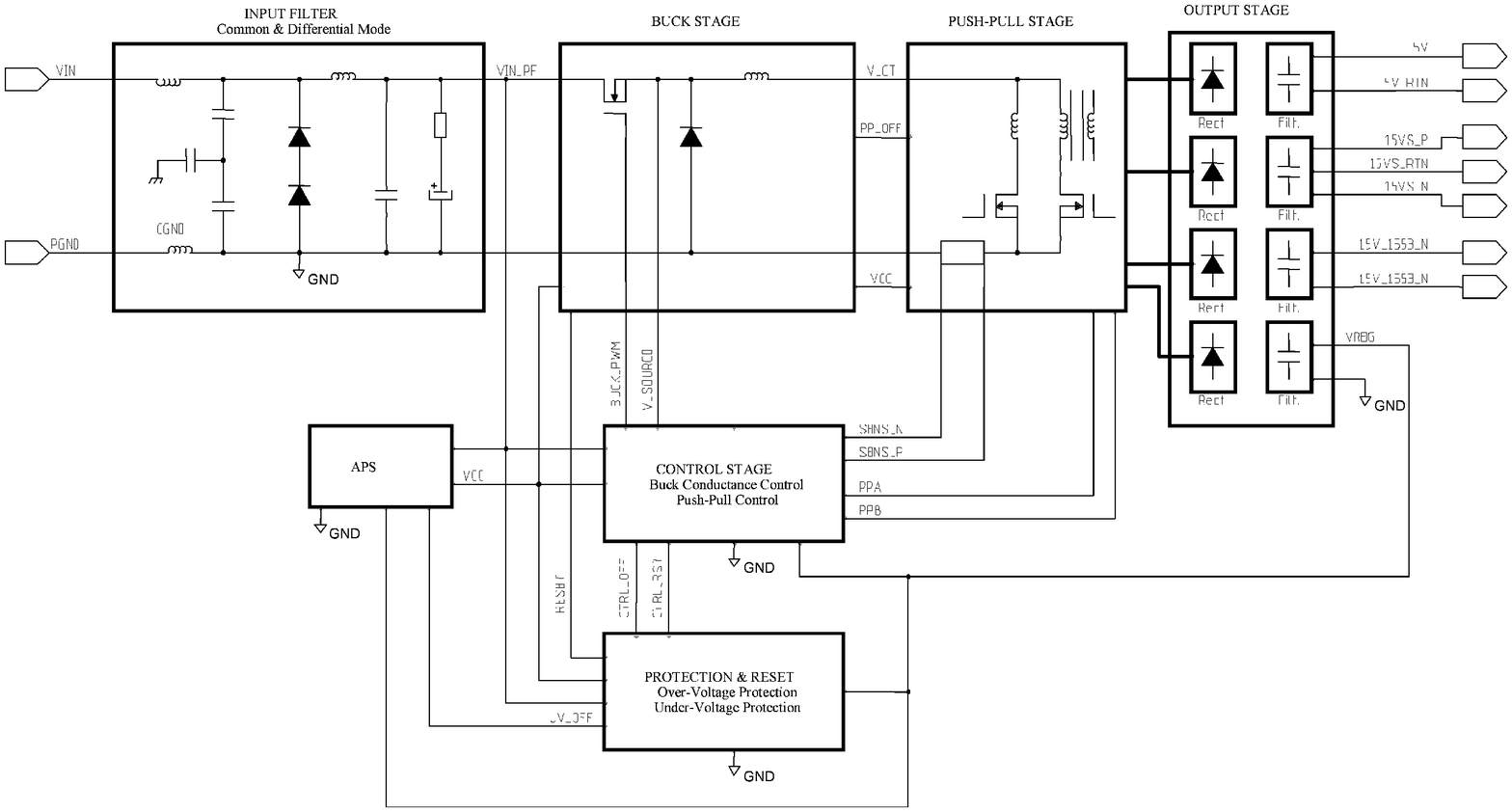}
  \caption{DC/DC converter block diagram. }
      \label{fig:1-7}
\end{figure*}

Figure~\ref{fig:1-7} is a block diagram of the DC/DC Converter which uses the
buck fed topology commonly used in space programs. The converter operation
starts once the power bus is present at the input. There is no command to turn
On or Off the converter.  The ``Buck stage'' provides the voltage regulation
at the output while the ``push-pull stage'' transforms the voltage to the
output levels. The converter has both common mode and differential mode filters
to reduce the emissions onto the main bus as shown in the block diagram.

Once the PDUt power bus is available, the converter turns On in a smooth manner
because the inrush current is limited by means of an external latching current
limiter in the PDU to avoid the peak current due to the charging of the
differential filter. Once the input filter is fully charged the converter starts
operating. This is achieved by means of an under-voltage detector that does
not allow the operation of the converter until the filter is completely
charged. This is necessary to guarantee that the converter control electronics
is supplied at the correct voltage and that the converter starts in a well known
and safe mode. The DC/DC converter incorporates over-voltage protection.
Over-current protection is implemented on the primary side by means of the buck
stage in case of a short circuit or overload on the secondary side.

\subsection{Grounding concept}

\begin{figure*}
\centering
\includegraphics[width=0.85\textwidth]{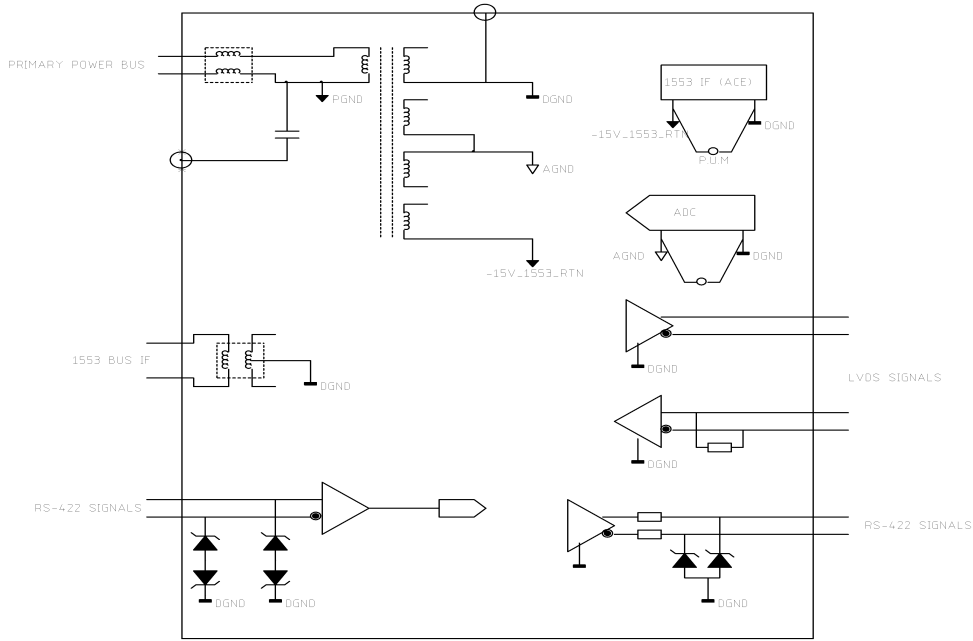}
  \caption{REBA Grounding diagram. }
      \label{fig:1-8}
\end{figure*}

The grounding of the REBA is via a single secondary return star point connected
to equipment chassis in a single point. Primary return is isolated from chassis
and from this secondary common return. Detailed grounding diagrams are shown in
Figure~\ref{fig:1-8}.  The different boards, including the mother board, include
ground planes for the different secondary returns. 

\subsection{Redundancy}

The whole of LFI was analysed during the early design phase for the effects of
failures, and a redundancy approach was adopted to reduce the possibility of
single point failures causing total loss of the instrument. As the REBA itself
would inherently be a point of single point failure unless very complicated
internal redundancy solutions were adopted that would themselves tend to
compromise the overall reliability of the unit, a simple approach was adopted of
replicating the unit for use in cold redundancy and thus also implementing cold
redundant interfaces to both the service module of the spacecraft (which
themselves were further redundant at the individual unit level) and the rest of
the instrument. In addition the failure analysis of the REBA itself was used to
make internal design choices and thus where reasonable mitigate single point
failures. For example margins were applied to the processor memories to allow
for the loss of single memory chips over the life of the unit in orbit. All
component choices in the REBA were tightly controlled by the Planck Product
Assurance Program under the product assurance rules of ESA and also burn-in and
ESA derating requirements were applied to limit the possibility of the most
likely cause of unit failure which would be in single components.

\section{Mechanical and thermal design description}

\begin{figure}
\centering
\includegraphics[width=0.85\columnwidth]{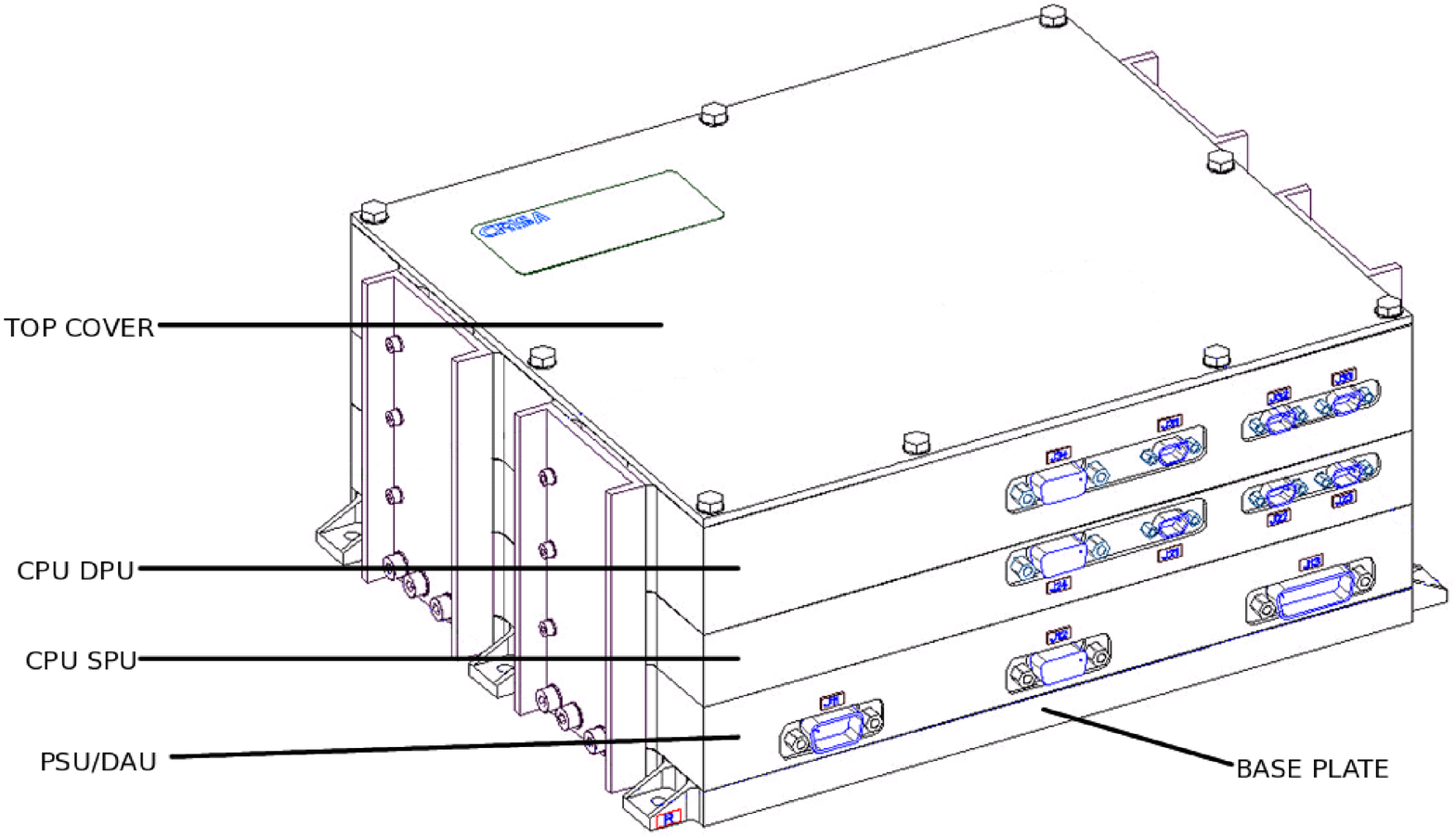}
  \caption{REBA BOX outline. The names indicate the modules distribution. }
\label{fig:1-9}
\end{figure}

The REBA is an assembly of three modules plus base plate with motherboard and a
top cover, distributed as is shown in Figure~\ref{fig:1-9}. In addition to these
modules the REBA has one independent motherboard, called the auxiliary board.

The principal objective of the thermal design was to avoid hot spots and
high thermal gradients between the different parts of the unit.  The PCB's are
coated with varnish to ensure good radiative coupling between boards and this
also contributes to the homogeneity of board temperatures. The PCB design
includes thick copper planes and the preferred location on stiffener and/or wide
copper tracks of components with high power dissipation. Externally, the box
parts are painted in mat black with an emissivity of 0.9.

The housing concept is an assembly of machined parts assembled by screws.
The housing modules are mounted horizontally, stacked one on the top of the other.
The REBA has two identical CPU boards, DPU (Data Processing Unit) and SPU
(Signal Processing Unit), with some differences in terms of size of memory
banks and external interfaces. The DPU Module has a higher stiffener than the
SPU Module because it houses the auxiliary PCB.

The I/O connectors and layout of the modules have been optimised in order to
minimise the height of the box. MHD connectors serve as interfaces for
electrical interconnections between modules through the motherboards.

\begin{figure}
\centering
\includegraphics[width=0.85\columnwidth]{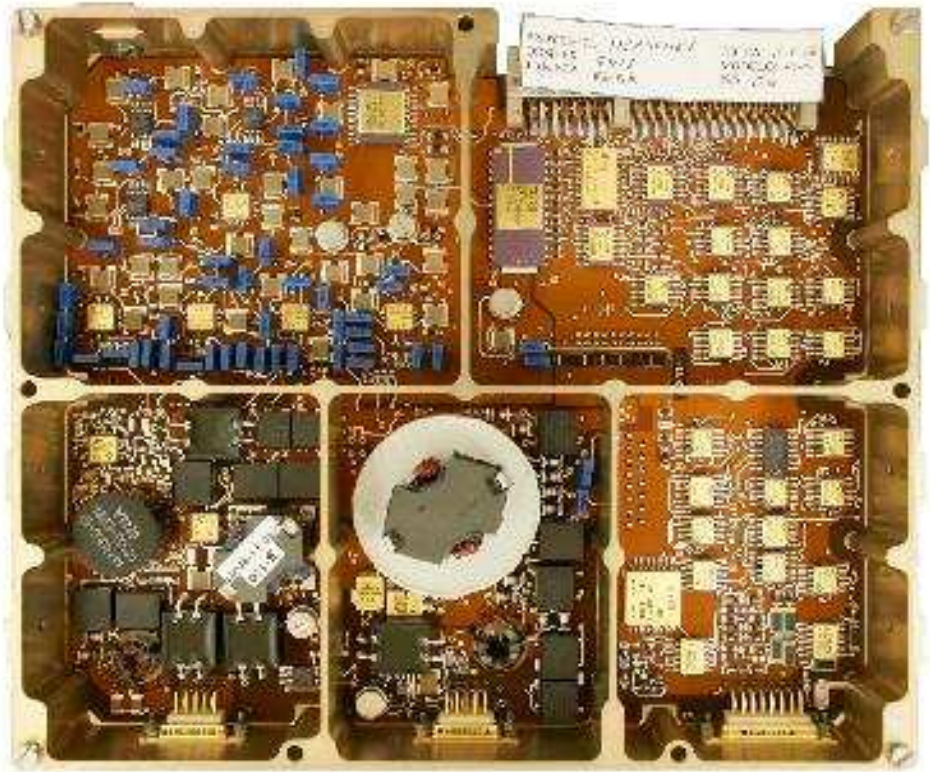}
  \caption{PSU and DAU module. }
      \label{fig:1-10}
\end{figure}

\begin{figure}
\centering
\includegraphics[width=0.85\columnwidth]{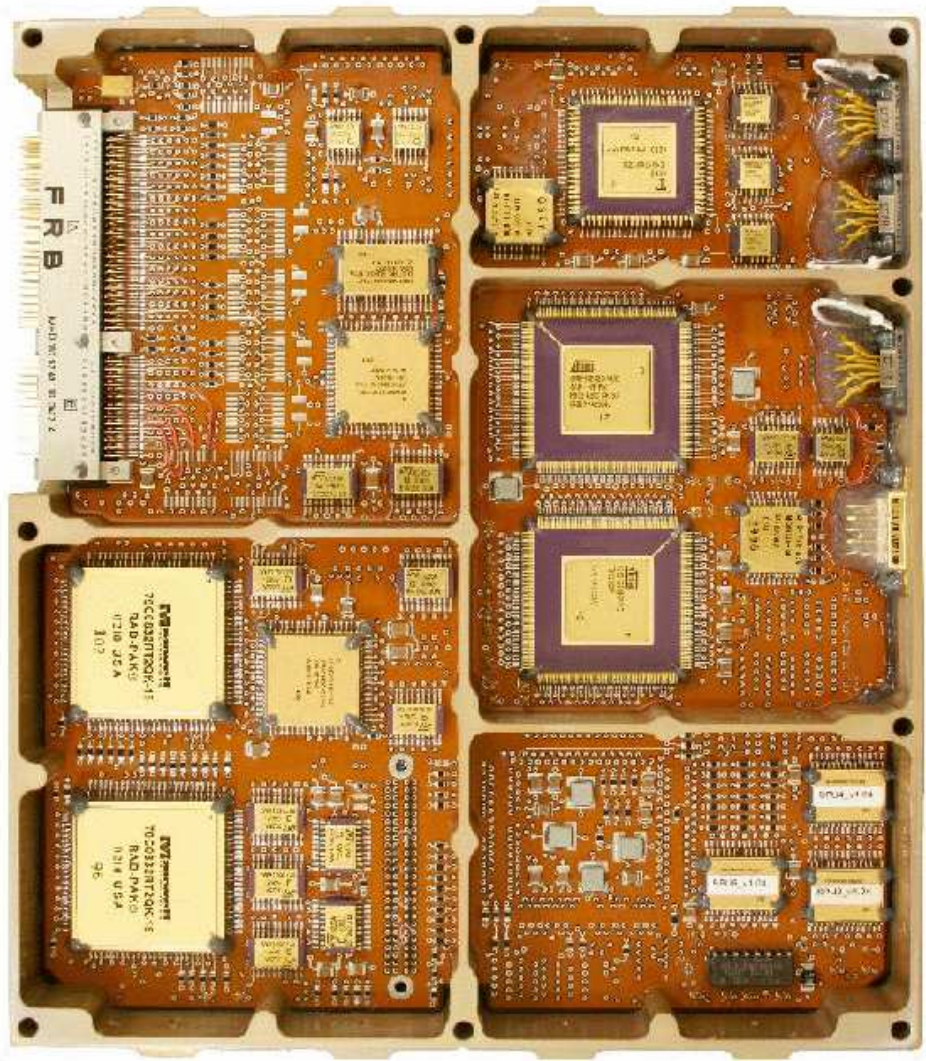}
  \caption{CPU SPU module. }
      \label{fig:1-11}
\end{figure}

\begin{figure}
\centering
\includegraphics[width=0.85\columnwidth]{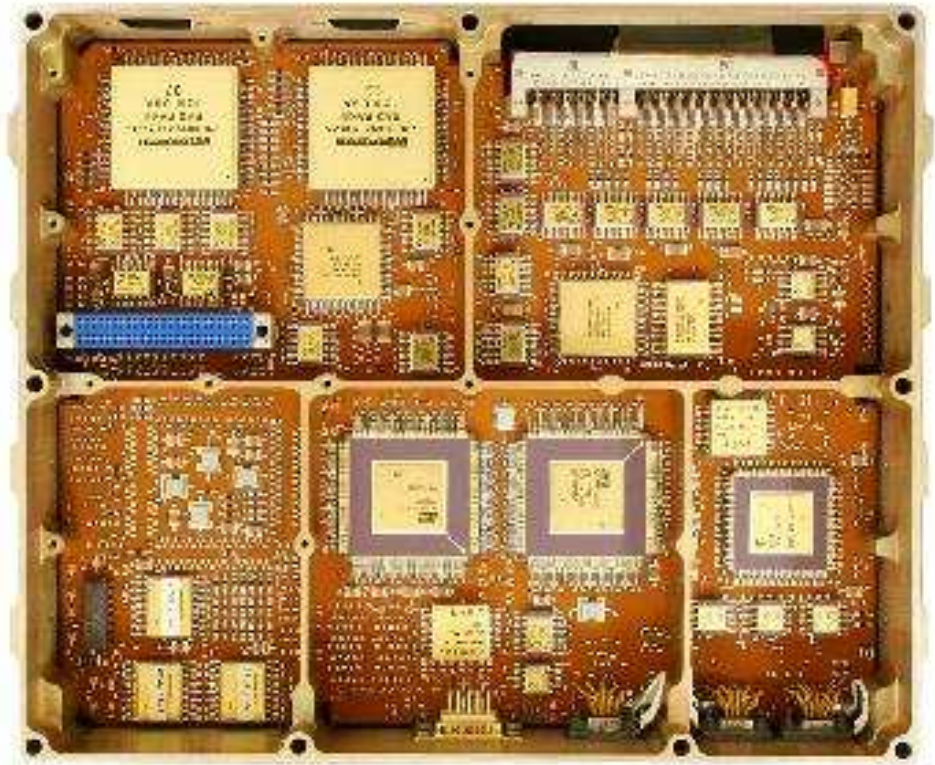}
  \caption{CPU DPU module. }
      \label{fig:1-12}
\end{figure}

\begin{figure}
\centering
\includegraphics[width=0.85\columnwidth]{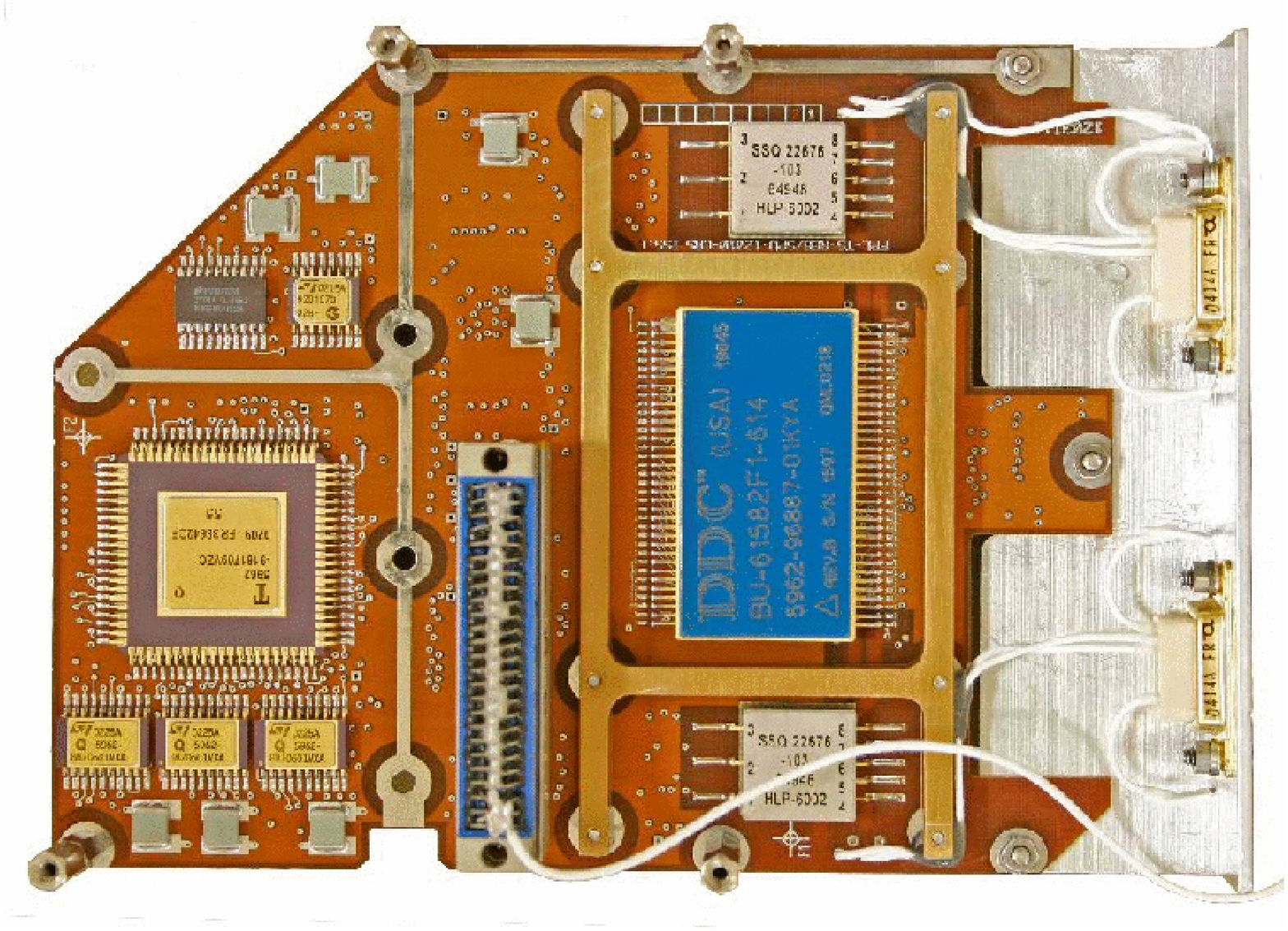}
  \caption{Auxiliary board. }
      \label{fig:1-13}
\end{figure}

The Base Plate is a machined plate having a flat contact surface to the
platform. This item has a pattern of mounting feet, total 6 (M4), located along
the housing perimeter and machined from a single block providing a continuous
flat bottom surface. The Top Cover is a 2 mm thick aluminium sheet.
The modules that form also the sides of the boxes have lateral reinforcements to
provide stiffness and thermal conductive coupling between the modules and the base
plate. The boxes are assembled by means of screws, passing through the Top Cover
and through the stiffeners to the base plate threaded inserts. The items of the
housing are machined parts made of aluminium alloy, having through holes or
threaded inserts along their edges to allow easy assembly, minimising the number
of mechanical parts. Figures~\ref{fig:1-10}, \ref{fig:1-11}, \ref{fig:1-12} and
\ref{fig:1-13} show a view of the modules.


The PCB is used for mounting components as well as printing wiring
interconnections between components. The components selected are surface
mounted (SMT) whenever possible to minimise mass and volume. The components were
selected for insertion mounting where high thermal dissipation was necessary
through the stiffeners or the SMT versions were not available. The PCB is made of
a resin polyimide and fibreglass multilayer laminate. The required surface to
allocate the electronic components is optimised in order to have a minimum number
of internal connections.

The stiffener frames have several functions: To provide adequate stiffness to
achieve natural frequencies on board above 140 Hz.  This is achieved by
machining them from a block of aluminium alloy with contours along the edges of
the boards, and inner ribs dividing the PCB areas in to smaller unsupported
sections. To provide PCB fixation by means of M2.5 screws in a sufficient number
to guarantee structural redundancy and thermal conductive coupling. To serve as
heat sinks for high dissipation components, such as power mosfets.  To provide
fixation for the I/O D*M and MDM connectors located on the front of the units.
Internal interconnections between modules is performed by a motherboard using
Medium High Density (MHD) connectors. The material used for the mechanical parts
of the box is aluminium 6082-T6 and its surface treatment is chromate and Black
paint.

\section{Design, development and AIV}

The following models were manufactured and tested at various levels. Two
identical engineering models: EM and AVM, the engineering qualification model:
EQM and the flight model: FM (both Nominal and Redundant). The REBA EM and AVM
models are flight model representative of both the electronic circuitry
interfacing the S/C and support the LFI flight software. Commercial electronic
parts were used with the same technology and by the same manufacture for the
circuits directly interfacing with the S/C subsystem as in the flight H/W.
The REBA-EQM is representative of the FM with the following exception: the EQM
electronic components are "extended range" type only; nevertheless, they have
been procured from the same supplier as that of the FM components and have been
produced with the same FM technology. The REBA-FM has full flight standard
components and was verified by formal functional and environmental acceptance
tests. In order to save costs flight spares (FS) boards have been produced for
replacement of failed or damaged equipment at the integration or launch sites
if it had been necessary.

The development of the REBA was split into four main phases: Preliminary Design
Phase, Detailed Design Phase, Flight Design Qualification Phase and Flight
Acceptance Phase. During the Preliminary and Detailed design Phases the
equipment concept was confirmed, the interfaces were defined, and the
requirements and the design were frozen. The Preliminary and Critical Design
Reviews were also held. The EM and AVM model were produced during this
stage. During the qualification Phase the flight baseline design configuration,
the preparation of the plans and procedures for QM and FM testing, and the
manufacturing, assembly, integration and qualification tests of the REBA QM were
performed. This Phase was completed with the Qualification Review. The Flight
Acceptance Phase was devoted to the manufacturing, assembly, integration and
testing of the REBA, nominal and redundant, flight equipment.  The production of
these models concluded with the corresponding Flight Delivery Review to allow
integration of the REBA in the flight model of LFI.

\begin{figure}
\centering
\includegraphics[width=0.85\columnwidth]{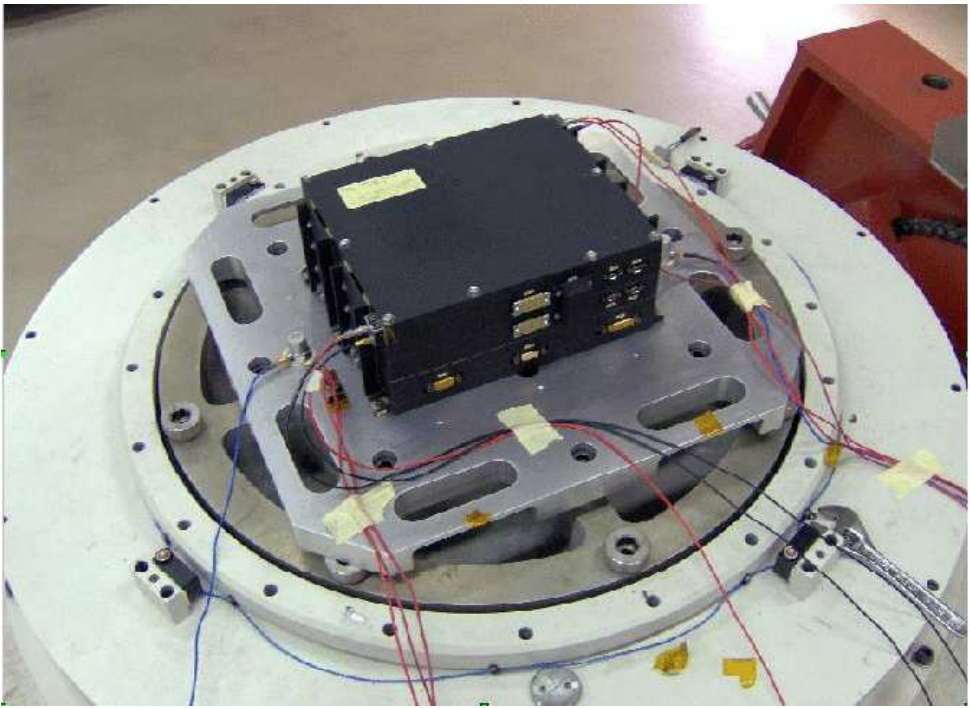}
  \caption{REBA Vibration tests. }
      \label{fig:1-14}
\end{figure}

Full self-procurement was established for the EEE parts for the EM and
AVM. Components for the EQM and FMs were procured by a Central Parts Procurement
Agency provided by ESA. Other specific components and materials such as ASICS,
magnetic components, mechanical parts, etc., were procured by CRISA. All the
assembly activities of electronic modules including equipment integration were
performed by CRISA. A dedicated Product Assurance Plan was prepared to cover the
quality assurance aspects related to the engineering, procurement, manufacturing
and verification activities involved in the development of the flight models.

The overall verification program consisted of two parts, the qualification
program and the acceptance program. Qualification tests are performed on the EQM
model, and acceptance tests on the FM model. The different models were submitted
to test in order to verify their characteristics and behaviour in the following
areas: physical, electrical, functional, mechanical, thermal vacuum and
EMC. Table~\ref{table:3} provides a review of these tests performed on the
qualification and flight models in a test matrix form.

\begin{table}
\caption{Verification matrix. The legend for the second and third columns is: (T) Tested; (Q) Qualification level and duration; 
(A) Acceptance level and duration; (R) Reduced (only in case of NC in QM tests).  }
\label{table:3}
\centering                          
\begin{tabular}{l c c}       
\hline
Test	& QM	& FM\\
\hline
Inspection Tests	& T	& T \\
Dimensions, mass, flatness \& Roughness	& T	& T \\
Centre of Gravity, Moment of Inertia	& T	& T \\
Grounding, Electrical and Functional	& T	& T \\
Shock Load Test	& Q  &   \\	
Sine Vibration	& Q	& A \\
Random Vibration	& Q	& A \\
Thermal Vacuum	& Q	& A \\
Conducted Emissions and Susceptibility	&T	&T \\
Radiated Emissions and Susceptibility	&T	&R \\
ESD &	T & \\
\hline
\end{tabular}
\end{table}


\section{REBA On-board software overview}

All the software of the LFI instrument is implemented in the Radiometric
Electronic Box Assembly (REBA). The LFI On-board software is divided into the
following software products:
\begin{itemize}
\item REBA DPU Start-Up Software: Developed by CRISA. It performs the
  initialization of the DPU and allows the upload into memory of the DPU
  Application Software (DPU\_ASW) which is stored in the DPU EEPROM. It also
  performs some hardware self tests to confirm the correct status of the unit.
\item REBA SPU Start-Up Software: Developed by CRISA. It performs the
  initialization of the SPU and allows the upload into memory of the SPU
  Application Software (SPU\_ASW) which is stored in the DPU EEPROM. It also
  performs some hardware self tests to confirm the correct estate of the unit.
\item REBA Low Level Software Drivers: Developed by CRISA. These provide a set
  of functions which allows the application software to access the different
  parts of the hardware.
\item REBA Application Software: Developed by IAC. It performs the complete
  operation of the LFI instrument. It is split in two applications running in
  the two CPU's inside the REBA, namely, DPU application Software (DPU\_ASW) and
  SPU application Software (SPU\_ASW).
\item Compressor and Decompressor Software: Developed by IAC, the compressor is
  implemented as a function of the SPU\_ASW. Its development was performed as a
  separate product and then, integrated in to the SPU\_ASW.
\end{itemize}

\subsection{REBA Low-level Software}

The REBA low level software consists of the Start-up Software (SUSW) and the
Low-Level software drivers (LLSW\_DRV). The DPU and SPU SUSW perform the necessary
functions to boot the unit at power up, perform a health self-test and start the
ASW giving it the control. Both programs are stored and executed from the DPU
and SPU PROMs.

The DPU SUSW is in charge of performing the following tasks: Power up
initialization. Reset source discrimination. Hardware self-tests. Hardware
initialization/configuration. Perform command reception/execution loop,
receiving commands and sending responses through the 1553 interface. Transfer
control to ASW, upon reception of appropriate command. The SPU SUSW is in charge
of performing the following tasks: Power up initialization. Reset source
discrimination. Hardware self-tests. Hardware
initialization/configuration. Perform command reception/execution loop,
receiving commands and sending responses through the 1355 interface. Transfer
control to ASW, upon reception of appropriate command.

The Low-Level software drivers (LLSW\_DRV) package is a set of primitives
(source/object code) that provide easier access to functionalities of the
different DPU/SPU HW devices which are compiled/linked with the ASW. The
LLSW\_DRV includes the following drivers: MILSTD 1553 (ACE device). Watchdog.
OBT. Data Acquisition Unit. SMCS 1355. Interrupts. EDAC. EEPROM. General
definitions. Error codes. Hardware map. DSP 21020 driver. PSC driver. Software
reset function. Memory management. Both, LLSW\_DRV and SUSW are coded in C and
Assembler languages and are implemented using the Analogue Devices Software
Development Environment ADSP-21020 Family Development Tools, release 3.3 for
PC. The Architectural design was performed using HOOD/UML tools. Software
requirements verification was carried out by test, analysis or assessment
(review of the design). Unit testing was performed using the ATTOL unit tests
tool and for the software integration the Analogue Devices DSP 21020
Emulator. SUSW and LLSW\_DRV software packages have been developed according to
Planck LFI REBA/Herschel PACS SPU SW Quality Assurance Plan and comply with the
CRISA SW Configuration Management Plan.

\subsection{REBA Application Software}

\subsubsection{Functionalities}

\begin{figure*}
\centering
\includegraphics[width=0.85\textwidth]{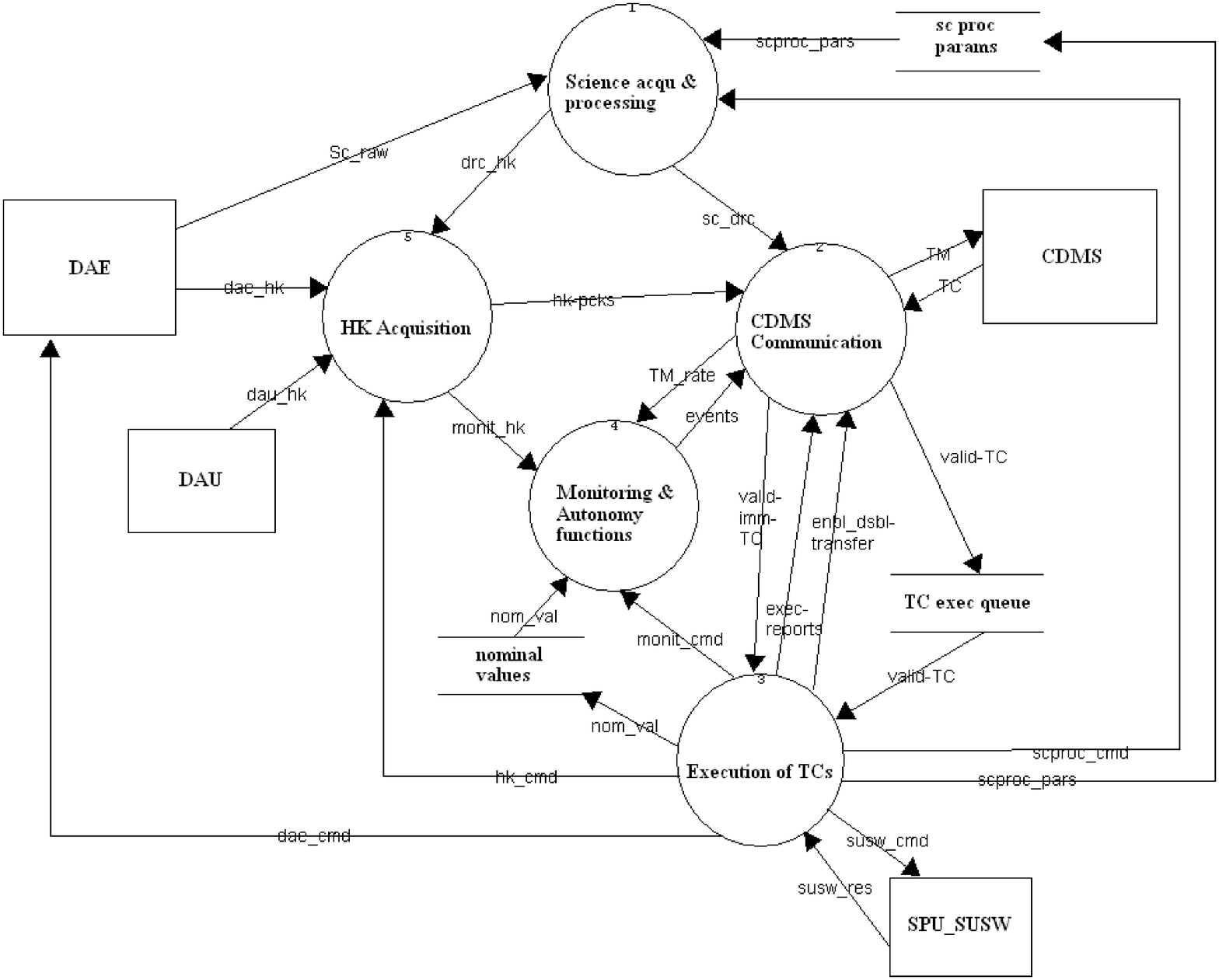}
  \caption{REBA\_ASW main functionalities. }
      \label{fig:1.15}
\end{figure*}

The Data Flow Diagram in Figure~\ref{fig:1.15} shows the main functions
performed by the REBA\_ASW.  The CDMS communication component is in charge
of packing and transferring all data produced in the instrument (Telemetry):
science data reduced and compressed (sc\_drc), housekeeping packets of the
whole instrument (hk\_pcks), events and execution reports (exec\_reports),
according to CDMS protocols (high and lower levels). This component also
receives and decodes every TC received from the CDMS according also to CDMS
protocols (high and lower levels) to produce valid instrument commands ready
to be executed, immediately (valid imm TC) or waiting in an execution queue
for eventual previous TCs execution (valid TC).

The main function of the REBA\_ASW is to receive from the Data Acquisition
Electronics (DAE) the raw science data (Sc\_raw) collected by the LFI
radiometers and reduce and compress these data (Sc\_drc) to within the allocated
bandwidth allowed to the instrument. The baseline for the science data processing
is to average, perform differences and compress the data of all the detectors,
nevertheless, the type of processing as well as the selection of the detectors
to be processed is modifiable by TC (scproc\_cmd). Some parameters needed for
the science processing can also be modified by TC (scproc\_pars).

The REBA\_ASW also collects housekeeping from the whole instrument. These data
are collected from the Data Acquisition Unit (DAU) inside the REBA (dau\_hk),
which supplies information on the REBA itself, and from the DAE (dae\_hk) which
supplies information about the rest of LFI instrument. Other housekeeping is
produced by the REBA\_ASW itself, as is the case for the processing of the science
data (drc\_hk) as well as other parameters (not included in the diagram for
simplicity). This function is performed by the HK acquisition component. The
operator may also require by TC (hk\_cmd) special HK information for
diagnostics and debugging purposes. The housekeeping collected are sent to the
CDMS as specific types of TM packets (hk\_pcks) and some of these are also
checked by the REBA\_ASW to control the health of the instrument (monit\_hk).

This automatic checking is performed by the ``Monitoring \& Autonomy functions
component'', which takes care of the CPU load in the SPU, temperature of the
focal plane; science TM rate and instrument internal communications links. When
the chosen values exceed certain nominal values (nom\_val), an alarm is sent to
the CDMS (events) and an autonomy function is activated to maintain the system in
safe conditions. It is also possible to command (monit\_cmd) this component for
testing and debugging purposes.

Finally, the other main component of the REBA\_ASW is the management of the TCs
received from the CDMS, Execution of TCs component. It controls the correct
execution of the incoming TCs (valid\_TC and valid\_imm\_TC) which operate the
instrument. It only allows the execution of TCs under certain conditions of the
system. It sends the TC to the appropriate subsystem or component when needed,
checks the completion execution and notifies the CDMS about the results
(exec\_reports). The transmission of TM packets to the CDMS may also be modified
by TC (enbl\_dsbl\_transfer) for testing and debugging purposes allowing the
enabling and disabling of the transmission of different telemetry packets.

Detailed functional requirements as well as all software requirements are
defined in \cite{1},\cite{2}, and \cite{3} complete the specification of REBA
Application software which has been developed following the ESA software
engineering standards for small projects BSSC(96) 2.

\subsubsection{Architecture}

The REBA\_ASW is composed of two applications programs running on the two CPU's
of REBA, DPU application software (DPU\_ASW) and SPU applications software
(SPU\_ASW), respectively.

Both applications were developed in C language and are executed under the real
time operating system Virtuoso, from Eonic Systems. Some specific functions,
critical in time, have been implemented in assembler.

To cope with the functionalities described in the previous paragraph, REBA\_ASW
has been built by the use of different Virtuoso objects. It is composed of a set
of Virtuoso tasks which interface among them by the use of some Virtuoso Fifos,
memory maps as well as global variables. Problems of racing and concurrency have
been solved by an appropriate definition of a priority scheme of tasks as well
as the use of Virtuoso semaphores and resources. Virtuoso timers have also been
used to implement some cyclic activities. In Appendix~\ref{app2}, we list and
describe the most important tasks of the REBA\_ASW.

\subsubsection{Verification and Validation}

A total of 91 test procedures were defined and executed to ensure that the
REBA\_ASW fulfils its software requirements. The validation was performed
against the Software Specification Document, \cite{1}. Validation is completely
defined in \cite{4}.

The acceptance tests consist of a set of validation procedures in a specific
order. These acceptance tests have been executed for each REBA hardware model.

Unit tests have been executed at function level to check that each function
executes according to its detailed design. Different tests cases for each
function have been executed to guarantee a 100\% statement coverage. More
details about unit tests can be found in \cite{5}.

\subsubsection{Metrics}

Table~\ref{table:metrics} provides metrics which indicate the maturity of the REBA\_ASW.

\begin{table*}
\caption{Metrics. Legends in columns mean: (a) C Source modules. Number of C source files as well as include files;
(b) C functions. Number of functions implemented in C language;
(c) Assembler functions. Number of functions implemented in assembler;
(d) Code lines: Number of lines, not blank and not commented;
(e) Comment lines. Number of comment lines explaining the source code;
(f) Nesting Max: An average over the total functions of the maximum statement nesting level in the function (0 for sequential code);
and (g) McCabe: An average over the total functions of the McCabe's cyclomatic complexity, i.e. an average per function of the number of linearly independent control paths. }
\label{table:metrics}
\scriptsize
\centering                          
\begin{tabular}{l c c c c c c c}       
\hline
 & Source Modules &	C functions &	Assembler functions &	Code lines&	Comment lines &	Nesting Max &	McCabe\\
\hline
DPU\_ASW&	41	&264	&4	&10100	&14091	&2.53	&5.44 \\
SPU\_ASW& 	36	&139	&3	&4679	&7341	&2.09	&4.13 \\
Compressor& 	2	&10	        &0	&224	&	        &1.4	&3.7 \\
Decompressor&2	&9    	&0	&242        &		&2.22	&5.22 \\
\hline
\end{tabular}
\end{table*}


\section{The SPU Compressor Module}

\subsection{Compressor requirements}

The LFI scientific mission requires compressing the scientific data in order to
make maximum use of the satellite to ground communication bandwidth. IAC
developed the compressor function which has been integrated in the SPU
application software.  A decompression function was also developed and delivered
to the Data Processing Center (DPC) in order to be integrated in to the ground
application software.

The SPU compressor component is particularly complex and, therefore, a complete
set of formal software documentation has been generated and delivered
specifically for both compressor and decompressor functions, consisting of a
software specification document (\cite{6}), which includes SW requirements and
architecture design, a software verification and validation plan document
(\cite{7}) and its corresponding validation plan report (\cite{8}), and a unit
test plan and unit test report documents.

The compressor input signal is expected to be dominated by white gaussian
noise, with information content or entropy for 30~GHz channels of $H \approx
5.16$~bits, which allows for a maximum compression ratio of $Cr_{op} \approx
3.10$, and the specified minimum compression ratio for such a signal is about
$Cr_{req} \approx 2.5$. In addition, the compressor is required to be lossless and
adaptive, which means recovering the exact input signal after decompression, and
optimizing compression performance in spite of variations in the input signal
statistics, respectively. All these requirements, together with the common
requirements of on board compressors of working in real time and packetizing
output data in independently decompressable packets, has led us to a careful
selection of the compression algorithm, and a great deal of work to adapt it to
the LFI compressor requirements.

Arithmetic coding is the method of choice for lossless adaptive entropy
compression on multisymbol alphabets because of its high speed, low storage
requirements and effectiveness of compression. The final SPU compressor is an
adaptation of the implementation of the arithmetic coder by A. Moffat, a widely
used non commercial version which has become a standard (\cite{9}).

\subsection{Compressor general architecture}

\begin{figure*}
\centering
\includegraphics[width=0.9\textwidth]{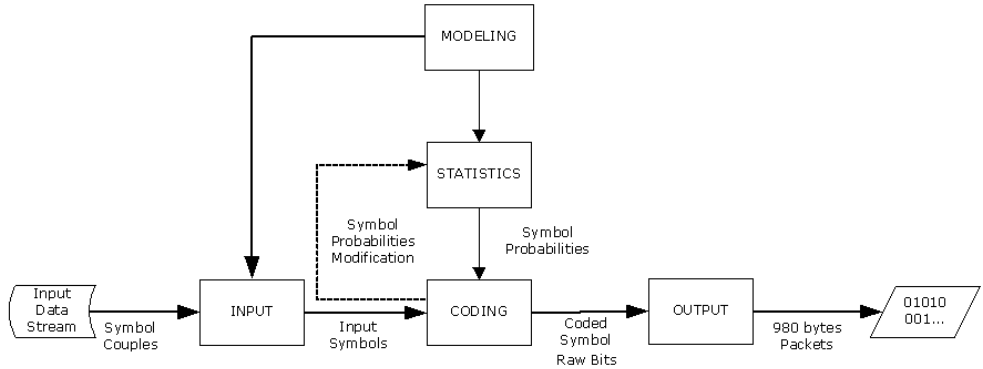}
  \caption{Modeling, statistics, coding, input and output modules. }
      \label{fig:1.16}
\end{figure*}

Adaptation of Moffats algorithm to LFI requirements has been viable thanks to
its modular architecture. Figure~\ref{fig:1.16} shows the modular architecture
of the SPU compression function (\cite{10}). This function is called by the SPU
application SW with the aim of generating a single compressed data set for a
single science TM packet. Each time the function is called, all the relevant
compression variables are set to the same initial values.

The coding module is the core of the compression function, and those options
that maximize the compression factor without penalizing execution speed have
been chosen. For example, multiplications and divisions are allowed, because
they are supported by the DSP as primitive operations, and permit a better
compression efficiency; and frugal bits mode is selected, so that coding
termination is made with as few nonambiguating bits as possible.

The modelling module informs the rest of the modules about the nature and
characteristics of the signal. The model is a 16 bit integer zero order adaptive
model, which means that 16 bits uncorrelated input samples are assumed, with an
unknown initial statistic.

The statistics module manages the Fenwick Tree structure that stores the symbols
cumulative frequencies. Accessing the structure takes $O(log_2(n))$ instructions
per symbol, where $n$ is the size of the alphabet, and so the cost is small even
when $n$ is very large (which is our case with $n = 2^{16}$ symbols). Initially,
the symbols occurrence frequencies are set to zero. In order to code a symbol
with a zero occurrence frequency, an escape symbol is coded followed by the raw
symbol. This allows a fast initialization that would otherwise prevent real
time operation. This also allows the statistics table to truly represent the
signal statistics right from the beginning of the compression operation with as
little increment in a symbol occurrence frequency as unity per new symbol, so
that no time consuming functions are needed to read just the statistics table
because of overflow.

Input and output modules read the input data and write the output data according
to the interface requirements with the SPU application software. The output
modules take care that the maximum allowed size of a compressed data set per TM
packet is not exceeded.

\subsection{Performance validation and results}

A formal SW validation has been conducted (see \cite{8} and \cite{7}) against
the compressor SW requirements specified in \cite{6}. Concerning  performance
on compression ratio and execution time the developed compressor is compliant.
For a Gaussian white noise signal with an optimal compression ratio of around
$Cr_{op} \approx 3$, the minimum achieved ratio is $Cr = 2.5$, and the maximum
execution time per generated packet is 46~ms.

\begin{figure*}
\centering
\includegraphics[width=0.9\textwidth]{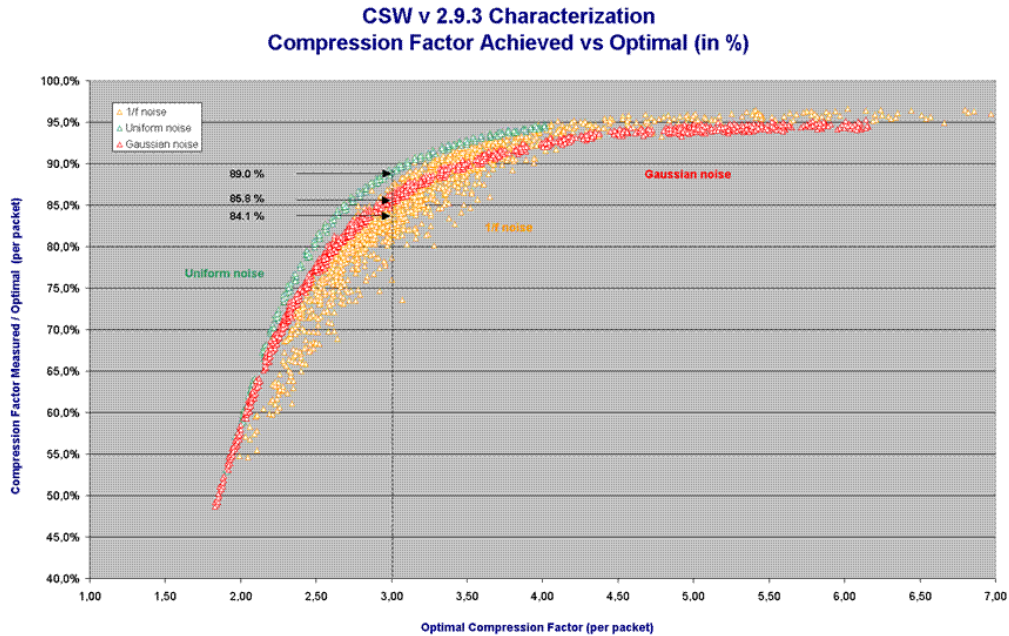}
  \caption{Achieved compression ratio per packet (in \%) vs. the optimal
    compression ratio per packet for Gaussian, uniform and $1/f$ noise
    distributions. }
\label{fig:1.17}
\end{figure*}

Also the compressor ratio has been characterised against noise with different
probability density functions (see \cite{11}). Figure~\ref{fig:1.17} shows the
percentage of achieved compression ratio versus the optimal, for Gaussian,
uniform and $1/f$ types of noises. We can see that the compressor behaves
slightly better for uniform noise than for the other two types. This is because
a uniform probability distribution is more robust against roundoff error due to
integer arithmetic divisions when calculating the symbol's probabilities.
However, for a Gaussian noise, which is the most representative of the actual
expected scientific input signal, the achieved compression ratio is
specification compliant.

\section{Conclusions}

The Radiometer Electronics Box Assembly (REBA) is the control and data
processing on board computer of the Low Frequency Instrument (LFI) of the Planck
mission (ESA), consisting in two separate identical units, which operate in cold
redundancy under power supply control of the Planck spacecraft.  Each REBA unit
is connected to the instrument DAE and to the spacecraft CDMU and PCDU.
REBA performs, among other functions, on-board command and data handling,
science data processing and compression, software storage and processing, and
control of the LFI. It has four functional units: DPU, SPU, DAU and PSU. It is
composed of two DSP boards and one board for the DAE interface function and the
DC/DC Converter.

We have manufactured and tested at various levels engineering models (EM and
AVM), the engineering qualification model EQM and the flight model FM (both
nominal and redundant). Specific software was developed to carry out the
start-up of the DPU and SPU, to perform the complete operation of the LFI
instrument, and to carry out on-board data compression in order to make maximum
use of the satellite to ground communication bandwidth. The compressor function
was integrated in the SPU application software.  The performance on compression
ratio and execution time was found in agreement with the specifications.

After launch of the Planck satellite on May 14th 2009, the REBA hardware and
software performs well and fulfil successfully all the specifications for flight
operations.

\acknowledgments 
Planck is a project of the European Space Agency with
instruments funded by ESA member states, and with special contributions from
Denmark and NASA (USA). The Planck-LFI project is developed by an International
Consortium lead by Italy and involving Canada, Finland, Germany, Norway, Spain,
Switzerland, UK, USA. The development of REBA has been funded by the Spanish
Ministry of Science through the National I+D+i Plan with several research grants
awarded in the period 1998-2008. We would like to acknowledge the very
significant contribution of the company CRISA to this project. We also thank
INTA for providing access to their facilities. We are indebted to the IAC
technical, scientific and administrative staff for their support. The authors
also wish to acknowledge the funding by ASI to the Italian LFI Program.

\newpage

\appendix

\section{Abbreviations and acronyms}
\label{app1}

\begin{tabular}{l l}
ASIC	& Application Specific Integrated Component \\
ASW		& Application Software \\
AVM		& Avionic Model \\
CDMS	& Central Data Management System \\
CDMU	& Central Data Management Unit \\
DAE		& Data acquisition Electronics \\
DAU		& Data Acquisition Unit \\
DMB		& Data Memory bus \\
DMPSC	& Data Memory Processor Support Chip \\
DPU		& Data Processing Unit \\
DSP		& Digital Signal Processor \\
EDAC            & Error Detection and Correction \\
EQM		& Engineering Qualification Model \\
ESA		& European Space Agency \\
ESD		& Electro Static Discharge \\
FIFO		& First-In First-Out \\
FM		& Flight Model \\
FS		& Flight Spare \\
Gxxxx	& 230 xxxx \\
HW		& Hardware \\
IO		& Input/Output \\
Kxxxx	& 210 xxxx \\
LFI		& Low Frequency Instrument \\
LLSW	& Low Level Software \\
LLSW\_DRV	& Low Level Software Drivers \\
Mxxxx		& 220 xxxx \\
PCDU	& Power Control Distribution Unit \\
PCB		& Printed Circuit Board \\
PMB		& Program Memory Bus \\
PMPSC	& Program Memory Processor Support Chip \\
PROM	& Programmable Read-Only Memory \\
PSC		& Processor Support Chip \\
PSU		& Power Supply Unit \\
RAA		& Radiometer Array Assembly \\
RAM		& Random Access Memory \\
REBA	& Radiometer Electronics Box Assembly \\
ROM	& Read-Only Memory \\
S/C		& Spacecraft \\
SMCS	& Scalable Multichannel Communication Subsystem \\
SPU		& Signal Processing Unit \\
SUSW	& Start Up Software \\
SVM		& Service Module \\
SW		& Software \\
TC      & Telecommand \\
TM      & Telemetry \\
\end{tabular}

\section{DPU\_ASW and SPW\_ASW architectures}
\label{app2}

\begin{table}[h!]
\caption{DPU\_ASW Architecture: Tasks}
\begin{tabular}{|l|p{3.5in}|l|}
\hline
Task & Description & Priority \\
\hline
tDpuInit & This task is the execution entry point of the complete
application. It initializes all variables and Virtuoso objects (semaphores,
timers, etc.) and starts the rest of tasks. After this initialization it aborts
its own execution. & 8 \\
\hline
tCdmsDrv & Implements the transfer layer protocol to communicate with the
CDMS. This communication is based on an interrupt handler which is executed at
reception of two types of interruptions produced by the 1553 interface, that is,
one interruption at subframe level (1/64 seconds) and the second one at frame
level every second. The interruption at frame level is also used to perform the
synchronization of the REBA with the spacecraft time. It manages TCs received by
validating the TC received and executing those of immediate type and storing in
a FIFO and a memory map those which are queue TCs. These later are processed by
tTcExecution task. This task also interfaces with TM rate monitoring task by
storing in a global variable the total of delivered science TM packets. & 10 \\
\hline
tDpuDmpscDrv	& This task manages events and errors produced by the Processor
Support Chip (PSC) of Data Memory of SPU. These events are detected by
interruption. This includes the management of EDAC single and double failures in
data memory, watchdog and hardware errors on the 1553 interface. Single EDAC
failures are also corrected by this task. 	& 16 \\
\hline
tDpuPmpscDrv & This task manages events and errors produced by the Processor
Support Chip (PSC) of Program Memory of DPU. These events are detected by
interruption. This includes the management of EDAC single and double failures in
program memory as well as the 1 Hz signal which is used to perform the
synchronization of DAE. & 17 \\
\hline
\end{tabular}
\end{table}

\begin{tabular}{|l|p{3.5in}|l|}[h!]
tDpuSmcsDrv & This task implements the low level communications of the 3 links
of the 1355 interface. Link 1 is used for communications between DPU and SPU and
link2 and link3 is used to communicate DPU with DAE. While link 3 is used for
transmission of data between DPU and DAE, link 2 implements the called "control
by link", that is, as DAE has no software, the operation of the DAE SMCS chip
must be commanded through a link, so, DPU commands DAE SMCS to perform the
appropriate transmission and receptions through this link. Depending on the
information of the interruption, this task signals appropriate semaphores or set
values in a FIFO; these Virtuoso objects are then used to communicate to other
higher level tasks of communications. & 18 \\
\hline
tDpuRcvFromSpu & This task processes the different packets received from the
SPU, the discrimination of the reception is based on a FIFO. The reception is
performed in a double buffer in such way that while a packet is being processed
another one can be received. According to the type of packet it performs
different activities, such as, saving SPU HK information for further packing of
REBA HK packet; obtaining science diagnostics data for further science
diagnostic statistics generation; storing in a FIFO the SPU CPU load for
monitoring, etc. & 25 \\
\hline
tRebaHkAcq &	Based on a 1 second timer, it collects nominal as well as
diagnostic acquisition data for REBA HK packet. The complete diagnostic data are
collected for each sample period and, depending on the transmission enabled of
TM packets, this is sent or reduced to just send the nominal values, a smaller
packet. & 	30 \\
\hline
tDaeHkAcq &	Based on a 1 second timer, it collects and packs the 2 types of
DAE HK data, that is, fast HK (every 1 second) and slow HK (every 64
seconds). It also selects and stores in a FIFO the Focal plane temperature (Fpt)
values for further monitoring analysis.  &  	31 \\
\hline
tFptMonit &	Implements the monitoring of the Focal Plane by checking the
values stored in a FIFO. After comparison with a threshold
value it also sends an event report and executes an autonomous function which
commands DAE to switch off the 4 DC/DC converters of the Front End Unit.
& 35 \\
\hline
\end{tabular}

\begin{tabular}{|l|p{3.5in}|l|}[h!]
tCpuMonit &	Implements the monitoring of the SPU CPU load by checking the
values stored in a FIFO. When limit is exceeded it sends an event report.
& 36 \\
\hline
tTmrMonit &	Based on a timer, whose period is configurable by the user, the
total of science TM delivered is checked. When the TM rate exceeds the expected
values an autonomous function to disable the processing in some detectors is
executed. The disabling is based on a table which can be configured by the
user. 	& 38 \\
\hline
tDpuEventMngr & 	All errors produced in the DPU system are managed by
this task. It processes all errors from a FIFO which is filled by all tasks when
errors are produced. This task also distinguishes when the error must be sent as
event reports. 	& 40 \\
\hline
tTcExecution	& Obtain from a FIFO the queued TCs pending of execution. These
TCs are already validated. The FIFO contains the storing information referred to
a memory map where the complete TCs are stored. Once the TC is executed, it
sends the corresponding execution report, success or failure. Synchronization of
both REBA and DAE is performed by the coordination of this task with tCdmsDrv
and tDpuPmpscDrv tasks. See later.	& 45 \\
\hline
tDpuMemoryScrub	& Based on a 24 hours timer, it reads all memories of DPU. This
allows to correct eventual single EDAC errors, decreasing, on this way, the risk
of double EDAC failures. 	& 60 \\
\hline
\end{tabular}

\vspace{1cm}
An important aspect to mention of the tTcExecution task is the synchronization
of REBA and DAE to the spacecraft time. REBA synchronization is carried out by
the use of global variable and a semaphore. TcExecution task set a flag to
indicate the synchronization process to tCdmsDrv task. When this flag is
asserted tCdmsDrv uses the one second interruption of 1553 interface to set the
REBA OBT according to the time received, a semaphore indicates to the
tTcExecution task that the synchronization has finished. A similar process to
verify the REBA time is followed. DAE synchronization process is similar; in
this case, tDpuPmpscDrv task guarantees that the synchronized time is passed to
DAE in advance to a 1~Hz interruption which is used by DAE to latch its time.

\newpage

\begin{table}[h!]
\caption{SPU\_ASW Architecture: Tasks}
\begin{tabular}{|l|p{3.5in}|l|}
\hline
Task & Description & Priority \\
\hline
tSpuInit & This task is the execution entry point of the complete
application. It initializes all variables and Virtuoso objects (semaphores,
timers, etc.) and starts the rest of tasks. After this initialization it aborts
its own execution. & 8 \\
\hline
tSpuDmpscDrv & This task manages events and errors produced by the Processor
Support Chip (PSC) of Data Memory of SPU. These events are detected by
interruption. This includes the management of EDAC single and double failures in
data memory, watchdog and the Data Ready signal coming from DAE which indicates
that a new set of science data has been stored in DAE DPRAM and it is ready for
being transferred to SPU. This is done by asserting a semaphore. & 16 \\
\hline
tSpuPmpscDrv & This task manages events and errors produced by the Processor
Support Chip (PSC) of Program Memory of SPU. These events are detected by
interruption. This includes the management of EDAC single and double failures in
program memory. & 17 \\
\hline
tSpuSmcsDrv & This task implements the low level communications of the 3 links
of the 1355 interface. Link 1 is used for communications between SPU and DPU and
link2 and link3 is used to communicate SPU with DAE. While link 3 is used for
transmission of science data from DAE to SPU, link 2 implements the called
"control by link", that is, as DAE has no software, the operation of the DAE
SMCS chip must be commanded through a link, so, SPU commands DAE SMCS to perform
the appropriate transmission through this link. Commands received from DPU are
stored in a FIFO which is processed later by the tCmdExec task. The
communications with DAE are managed by two semaphores. & 18 \\
\hline
\end{tabular}
\end{table}

\begin{tabular}{|l|p{3.5in}|l|}[h!]
tInitialScProc & This task starts its activity based on a semaphore which is
signaled when Data Ready signal, asserted by DAE, is detected. Performs initial
processing of science data including data acquisition from the DAE, data
validation, and depending on the processing type(s): data summing, differencing,
re-quantisation and offset. Data are stored in circular buffers implemented by a
Virtuoso memory map. It organize the science data separating by detectors and,
depending on the processing type, it stores raw data in circular buffers or
performs the sum of the data before store it. Communicates that a science packet
is ready to be compressed and/or packaged through a FIFO. & 25 \\
\hline
tSpuHkGen & Based on a 1 second timer, it collects SPU HK data and transmit this
data as a packet to DPU. & 30 \\
\hline
tSpuEventMngr & All errors produced in the SPU system are managed by this
task. It processes all errors from a FIFO which is filled by all tasks when
errors are produced. This task also distinguishes when the error must be sent as
event reports. & 40 \\
\hline
tScTxToDpu & Transmits a Science SPU packet to the DPU. These packets have been
previously stored in SPU DPRAM. A FIFO with memory addresses of Sc packets is
the interface with tFinalScProc task. & 43 \\
\hline
tCmdExec & It listens from link1 commands received from DPU and executes the
corresponding command. & 45 \\
\hline
tFinalScProc & Performs final processing of science data, including data
compression (depending on the processing type). It obtains packets ready to be
compressed and/or packaged. It interfaces with tInitialScProc through a FIFO. TM
packet is generated in a dedicated double buffer in the SPU DPRAM, this double
buffer allows to transmit a packet while new packet is being produced.  & 50 \\
\hline
tClbSwitch & Based on timer of 15 minutes, it enables each cycle a detector in
sum data processing type in one of 2 provided simultaneous processing groups. &
58 \\
\hline
tSpuMemoryScrub & Based on a 24 hours timer, it reads all memories of DPU. This
allows to correct eventual single EDAC errors, decreasing, on this way, the risk
of double EDAC failures. & 60 \\
\hline
\end{tabular}

\end{document}